\documentclass[10pt,twocolumn,twoside]{IEEEtran}
\IEEEoverridecommandlockouts

\usepackage{cite}
\usepackage{amsmath,amssymb,amsfonts,amsthm,bm}
\usepackage{algorithm}
\usepackage{algpseudocode}
\usepackage{graphicx}
\usepackage{textcomp}
\usepackage{balance}
\usepackage{xcolor}
\definecolor{revgreen}{HTML}{1B7F3B}

\usepackage[utf8]{inputenc}
\usepackage{booktabs}
\usepackage{multirow}
\usepackage{hyperref}
\usepackage{tikz}
\usepackage{float}
\usetikzlibrary{arrows.meta, positioning, shadows}
\usepackage[noEnd=false,indLines=true]{algpseudocodex}
\usepackage{enumitem}
\usepackage[subtle]{savetrees}
\usetikzlibrary{arrows.meta,positioning}
\usepackage{fvextra}
\usepackage{listings}
\usepackage{tcolorbox}
\tcbuselibrary{skins,breakable}

\newtcolorbox{PromptBox}[1][]{
  enhanced,
  breakable,
  colback=gray!5,
  colframe=gray!60!black,
  fontupper=\ttfamily\scriptsize,
  boxrule=0.5pt,
  arc=2pt,
  left=4pt,
  right=4pt,
  top=2pt,
  bottom=2pt,
  #1
}
\newtcolorbox{QueryBox}[1][]{
  enhanced,
  breakable,
  colback=black!3,
  colframe=black!40!black,
  fontupper=\small,
  title=#1,
  fonttitle=\bfseries,
  boxrule=0.5pt,
  arc=2pt,
  left=6pt,
  right=6pt,
  top=6pt,
  bottom=6pt,
  attach boxed title to top left={yshift=-2mm, xshift=2mm},
  coltitle=black,
  colbacktitle=white,
  boxed title style={boxrule=0.5pt, colframe=black!40!black}
}

\newcommand{\PC}{PowerDAG}
\newcommand{\AWF}{Annotated Query-Workflow Exemplars}
\newcommand{\AWFS}{Annotated Exemplars}
\newcommand{\awf}{annotated query-workflow exemplars}
\newcommand{\awfs}{annotated exemplars}

\usepackage{pifont}
\newcommand{\cmark}{\ding{51}}
\newcommand{\xmark}{\ding{55}}

\newcommand{\llm}[1]{\textsc{#1}}

\newsavebox{\algbox} % preamble

\def\BibTeX{{\rm B\kern-.05em{\sc i\kern-.025em b}\kern-.08em
    T\kern-.1667em\lower.7ex\hbox{E}\kern-.125emX}}
\usepackage{lipsum}

\renewcommand{\arraystretch}{0.92}

\title{\LARGE \bf 
PowerDAG:
Supervisory Agentic AI System for Automating Distribution Grid Analysis
}

\author{Emmanuel O. Badmus, Amritanshu Pandey%
\vspace{-0.5cm}
\thanks{Emmanuel O. Badmus and Amritanshu Pandey are with the Department of Electrical and Biomedical Engineering, University of Vermont, Burlington, VT, USA. Email: {\tt emmanuel.badmus@uvm.edu, amritanshu.pandey@uvm.edu}}
}

\begin{document}
\bstctlcite{BSTcontrol}
\allowdisplaybreaks

\maketitle

\begin{abstract}
Distribution grid analyses include tasks such as network information retrieval, power-flow analysis, hosting-capacity assessment, DER planning, and state estimation.
Completing these tasks often requires long-horizon, stateful workflows in which an engineer retrieves data, loads a feeder, runs simulations, evaluates results, and exports outputs.
The growing volume of these analyses is outpacing the limited engineering workforce, causing suboptimal outcomes and delays.
Large Language Model (LLM)-orchestrated agents can help, but they often struggle for two reasons: (i) they lack algorithms to determine the right context for an unseen grid task, and (ii) they cannot verify proposed actions against the environment state beforehand and instead rely on feedback after execution.
We propose \PC{}, an agentic artificial intelligence (AI) system that formalizes workflows as directed acyclic graphs (DAGs) and addresses current gaps in this formalism through two mechanisms, \textit{adaptive retrieval} and \textit{Just-in-Time supervision}.
To dynamically retrieve relevant context, it curates and ranks expert exemplars using an adaptive score-decay cutoff that matches the query complexity.
For supervision, it evaluates prerequisites before every tool call.
If an agent proposes an invalid action, the supervisor blocks execution, preserves the environment, and returns a corrective advisory.
% We evaluate \PC{} on a 200-record benchmark spanning ten task families.
% We evaluate \PC{} on 150 held-out queries from a 200-record expert-verified benchmark spanning 10 task families, 6 agent-system configurations, and 10 language models, resulting in 9,000 runs.
We evaluate \PC{} on 150 held-out queries from a 200-record expert-verified benchmark that covers 10 of the most commonly performed distribution-grid analyses, comparing 6 agentic systems across 10 LLMs for a total of 9,000 runs.
\PC{} reaches a success rate of 98.0\% with GPT-5.5, 97.3\% with Gemini 3.1 Pro, and 92.7\% with Qwen3.6-27B, improving success rates by 6 to 50 percentage points over baselines.
% \rev{Against standard agent baselines (ReAct, LangChain, and CrewAI), \PC{} improves the success rate by 6 to 50 percentage points, a statistically significant gain on nine of the ten models.}

\end{abstract}

\section{Introduction}
\label{sec:introduction}

\noindent The rapid growth of distributed energy resources (DERs), including rooftop photovoltaics, battery storage, and electric vehicles, has increased the number and complexity of distribution-grid studies that utilities need to conduct~\cite{bank2013analysis}.
These studies cover a broad set of planning and operation tasks, including power-flow analysis~\cite{kersting2018distribution}, dynamic hosting-capacity assessment~\cite{badmus2024anoca}, and related grid studies.
Increasingly, utilities, regulators, and policy groups must run many such analyses but often lack the engineering capacity to do so at scale~\cite{ewab2025opportunities}.
%{\color{blue}
% The challenge in distribution-grid analysis is not merely tool access.
% These studies involve stateful workflows across feeder models, time-series data, optimization routines, and solver pipelines.
% As a result, an agent can generate a syntactically valid tool sequence while still producing an invalid engineering result due to violated solver prerequisites, incorrect formulations, or stale environment state.
% This paper therefore targets physics-grounded workflow correctness: generating executable and auditable workflows that preserve dependency constraints and state consistency.
% }
This is partly because executing these analyses requires long-horizon, multi-step workflows across simulation and optimization tools (e.g., GridLAB-D~\cite{chassin2014gridlab}) and data pipelines (e.g., advanced metering infrastructure (AMI) databases), with correct tool ordering, argument binding, and environment state.
Rule-based automation engines can help, but they rely on fixed study templates and cannot accommodate the diversity of real-world queries, requiring continual manual curation~\cite{buchanan1984rule}.
LLMs offer an alternative by allowing engineers to describe an analysis in natural language and receive an orchestrated sequence of tool calls~\cite{schick2023toolformer}.

Traditional non-agentic approaches prompt the LLM for a full tool-call plan~\cite{patil2024gorilla,qin2023toolllm}.
However, the engineer stays in the loop, executing each call by hand and feeding the result back.
%At each step, the LLM recommends a tool call and arguments.
%The executor runs this on the current environment state, generating an observation (e.g., a success or error message) that is added to the context for the next iteration \cite{yao2022react}.
%These systems execute tool calls in a stateful environment, in which each call can read or modify shared objects that subsequent calls depend on.
Agentic systems close this gap. 
At each step, the agent proposes a tool call, the environment executes it, and the observation informs the next action~\cite{yao2022react}.
% This architecture enables fully automated orchestration, but introduces a challenge: because the LLM lacks direct access to environment state and decides actions probabilistically, it can generate tool calls that violate prerequisite dependencies, corrupting the execution state in ways that reactive feedback alone cannot prevent.
However, LLMs are not specifically pretrained on the tool-call logic of distribution-grid analyses, so they often produce incorrect tool selections, wrong call orderings, or invalid arguments~\cite{patil2024gorilla}.
Two primary strategies address this. Supervised fine-tuning (SFT) trains the model on domain data~\cite{qin2023toolllm}. While SFT can teach an LLM distribution-grid concepts, training it to orchestrate multi-step analyses using specific, changing tool sets remains difficult because it requires large domain-specific datasets and complete retraining whenever tools, models, or APIs change. In-context learning (ICL) avoids this by appending task-relevant context to the prompt without changing model weights~\cite{brown2020language,min2022rethinking}.
% ICL avoids these by conditioning the model on curated demonstrations, verified examples pairing queries with correct tool-call sequences, that encode procedural knowledge such as call ordering, state dependencies, and argument structures \cite{badmus2025powerchain,bhattaram2025geoflow}.

ICL avoids retraining by supplying the model with task-relevant context in the prompt~\cite{brown2020language}.
However, the context must encode cross-tool execution procedures.
General sources, such as papers and manuals, explain individual tools but rarely specify the data handoffs, call ordering, and state updates required across tools~\cite{min2022rethinking}.
% For example, documentation may explain how to run \textit{OpenDSS power flow} and how to query \textit{AMI data}, but it rarely specifies the end-to-end \textit{cross-tool} steps that \textit{bind AMI time series to feeder buses}, \textit{update the active model state}, and then \textit{run and validate} the solver and plots on that updated state.
%For example, GridLAB-D documentation \cite{chassin2014gridlab} explains how to \textit{run power flow} and how to use \textit{player/recorder} inputs, and separate database guides explain \textit{SQL/AMI queries}, but neither specifies the \textit{cross-tool} workflow that maps and aggregates AMI data into feeder load models and keeps downstream simulations consistent with the updated state.
For example, the GridLAB-D documentation and the AMI database guides each describe their respective APIs, but neither specifies how to map AMI measurements to feeder load models while maintaining downstream simulation consistency.

Recent work~\cite{BADMUS2027113555,bhattaram2025geoflow} addresses this by curating annotated workflow exemplars.
These are expert-verified query-workflow records that encode the correct call order, argument values, and state dependencies.
Although modern LLMs support long prompts, including all archived \awfs{} in every prompt increases token costs and can introduce irrelevant context, making orchestration less reliable~\cite{liu2024lost}.
% inclusion of irrelevant \awfs{} can confuse tool selection and increase token cost without benefit \cite{liu2024lost,sclar2023quantifying,TRIPATHI2025841}.
Prior work~\cite{BADMUS2027113555} mitigates this by retrieving the top-$k$ most similar \awfs{} per query.

% However, static top-k retrieval creates an inherent trade-off: small k omits critical procedural steps for complex queries, while large k includes irrelevant examples that inflate context and degrade tool selection. Moreover, similarity-based retrieval often matches query phrasing rather than procedural requirements. For instance, a query "check voltage violations" might retrieve examples about voltage analysis that assume a pre-solved power flow, but if the current execution state has no solution, the retrieved demonstrations guide the agent toward invalid tool calls.

\noindent \textbf{Problem:}
% Static top-$k$ selection of these examples creates a trade-off: small $k$ misses key steps for complex tasks, while large $k$ adds irrelevant examples that confuse tool choice and inflate context size \cite{liu2024lost}.
% Query-only retrieval often returns examples that match the query text but demonstrate the wrong procedure.
% To select, the system must adaptively choose examples based on both query similarity and example relevance.
Static top-$k$ retrieval has two drawbacks.
First, $k$ is a global constant.
Here, a small $k$ may miss the required tool steps for complex queries, while a large $k$ may include unrelated exemplars that confuse the LLM and inflate token costs~\cite{liu2024lost}.
Second, similarity-based retrieval matches query phrasing rather than procedural content. 
For instance, an exemplar may share keywords with the current query, but if it follows a different tool sequence, it can mislead the agent into skipping required steps.
% Consider an \awfs{} that codifies tool orchestration for a trivial query: \textit{plot bus voltage magnitudes from input network data for South Hero feeder in Vermont for March 2nd 2025.}.
% %performs only data loading and visualization without running a solver.
% When a user asks a previously unseen but semantically similar query: \textit{solve power flow and plot bus voltage magnitudes for South Hero feeder in Vermont for yesterday}, similarity-based retrieval may select the prior \awfs{} and induce the agent to skip the required power flow step and produce plots of unsolved or default values.

A second gap arises in the tool-call execution.
At each step, the agent calls an LLM to propose the next tool call, then executes it and reads the return message to decide whether to continue.
In stateful environments, tools can run successfully on stale or uninitialized objects without raising exceptions.
For example, if the agent skips a prerequisite \textit{network update} and calls the \textit{power flow solver} directly, the solver returns a success message while operating on the wrong state.
The agent accepts this as correct and proceeds.
Preventing such silent failures requires a technique that enforces these dependency constraints before each call, not one that reacts after the state is corrupted.

% {\color{blue}
% The contribution is a verifiable agentic workflow layer for distribution-grid studies, where exemplar retrieval targets procedural power-system workflows and supervision checks solver-backed state dependencies before execution.
% The framework combines adaptive exemplar retrieval, workflow-aware filtering, and external pre-execution supervision for stateful tools.
% Although portable in principle, the evaluation target here is workflow correctness for real distribution-grid analysis procedures.
% }

\noindent \textbf{Proposed Solution:}
To address both gaps, \PC{} makes three contributions:
\begin{itemize}
    \item \textbf{Adaptive, workflow-aware retrieval of \awfs{}.} Unlike fixed top-$k$ retrieval ranked on query-text similarity alone, we introduce a two-stage selector that adaptively accounts for query and workflow relevance. Stage 1 eliminates the fixed $k$ constant by fitting a two-segment score-decay model to adaptively select the \textit{most} similar candidate exemplars. Stage 2 then filters on procedural structure rather than query text, leveraging each exemplar's expert-verified tool-call trace to discard candidates whose tool sequences and prerequisite-state dependencies are inconsistent with the unseen query.
    \item \textbf{Just-in-Time (JIT) supervision.} We design a JIT supervisor, a deterministic guardrail, that checks, before each tool call, whether the environment satisfies its DAG-encoded prerequisites.
    If not, it blocks the call and returns a corrective advisory, leaving the environment unchanged.
    This stops the agent from running a solver on a stale or uninitialized state and accepting the result as correct.
    \item \textbf{Distribution-grid agentic benchmark.} We curate 200 expert-verified query-workflow records spanning ten distribution-grid task families, from single-step data lookups to multi-stage solver and optimization pipelines.
    We release them as the first benchmark for evaluating the correctness of agentic workflows in distribution-grid analyses.
\end{itemize}

%\noindent We validate \PC{} on 30 unseen distribution-grid queries and compare against state-of-the-art commercial and academic baselines.

% \noindent We validate our agentic AI system \PC{} on 30 distribution-grid tasks spanning four categories: solver-free data analysis, steady-state unbalanced three-phase power flow \cite{pandey2018robust}, dynamic hosting capacity \cite{badmus2024anoca,liu2022using,moring2023inexactness,yi2022fair}, and three-phase infeasibility analysis \cite{foster2022three,panthee2025solving}.

%First, we introduce adaptive retrieval of annotated examples that replaces the fixed top-$k$ with a cutoff selected from a similarity-score decay, and then filters candidates using both their query text and tool-call traces to provide procedurally relevant context for the agent.
%Second, we implement proactive prerequisite supervision that checks state-dependent tool requirements before each invocation, blocks calls that lack prerequisites to prevent state corruption, and returns a one-shot advisory directing the agent to the required setup steps. We evaluate our approach on a benchmark of 30 realistic distribution-grid tasks spanning four query families: (i) data-only queries that do not run a solver, (ii) steady-state unbalanced three-phase power flow, (iii) dynamic hosting capacity studies, and (iv) current infeasibility analysis.

\section{Related Works}
\label{sec:related_work}

\noindent This section reviews LLM tool orchestration, exemplar-conditioned orchestration, exemplar retrieval, and runtime supervision of tool execution.

\subsection{LLMs for Tool Orchestration}
\noindent Commercial LLM chat interfaces (e.g., ChatGPT, Claude, Gemini) concatenate user queries with historical conversation context and route them to their backend LLM (e.g., GPT-5.5, Claude Opus 4.5, Gemini 3.1 Pro) for response generation.
In standard chat mode, the LLM cannot run external distribution-grid simulators, so it returns generated code or guidance that the user must manually execute and validate~\cite{bonadia2023potential}.
Power-system applications include OpenDSS file generation~\cite{bonadia2023potential}, iterative script refinement~\cite{jia2025enhancing}, and result visualization~\cite{jin2024chatgrid}.
An alternative approach augments prompts with tool descriptions and asks the model to produce a complete tool-call sequence with bound arguments~\cite{patil2024gorilla,qin2023toolllm}.
%A similar approach is another approach that (iv) uses an LLM API to return single-pass tool calls for execution \cite{patil2024gorilla,qin2023toolllm}.
%{\color{red}These require manual tool calls.}
Such prompt-only systems still require the user to execute, inspect, and debug the resulting tool calls outside the chat interface.
Agentic frameworks close this gap by running an iterative loop, in which the LLM proposes one tool call, the environment executes it, and the observation informs the next step~\cite{schick2023toolformer,yao2022react}.
Agentic grid applications include GridMind for optimal power flow (OPF) automation~\cite{jin2025gridmind}, GridAgent for contingency analysis~\cite{zhang2025grid}, RePower for solver-driven planning~\cite{liu2025repower}, and X-GridAgent for executable multi-step workflows across power-system simulators~\cite{chen2025x}.
They also include PFAgent for automating power-flow studies with verification-driven refinement~\cite{she2026pfagent} and Grid-Orch for connecting LLMs to OpenDSS via Model Context Protocol (MCP) for distribution-grid simulation~\cite{liu2026grid}.
However, none of these systems explicitly enforce prerequisite dependencies.
State-dependency violations can therefore cause silent failures in stateful pipelines.
% {\color{blue}
% Plan-first and training-based methods are also related: OrchDAG models multi-turn tool use with synthetic plan DAGs and graph-structured reinforcement learning \cite{lu2025orchdag}. \PC{} instead targets real distribution-grid workflows with black-box tools and no policy training; the supervisor blocks prerequisite violations at execution time rather than internalizing dependencies from synthetic rewards.
% }
\subsection{Exemplar-Conditioned Orchestration}
\noindent Beyond prerequisite enforcement, agentic frameworks also struggle with domain-specific tool-call logic.
LLMs are not pretrained on these procedures, so without procedural guidance, they select the wrong tools, sequence calls incorrectly, and bind arguments to stale objects.
Recent systems condition decisions on retrieved exemplars of successful tool use, rather than relying on natural-language manuals alone~\cite{lewis2020retrieval,min2022rethinking}.
Tool-retrieval methods use iterative LLM-generated feedback to refine the selection of tools and exemplars~\cite{xu2024enhancing}.
Several agent systems store and retrieve prior trajectories as procedural memory to improve decisions in long-horizon tasks~\cite{wang2024agent}.
Other work builds reusable multi-step traces from past interactions and retrieves and refines them for new tasks~\cite{tan2025meta}.
GeoFlow explicitly names these traces \emph{workflows} and represents them as Activity-on-Vertex graphs with step-level tool objectives, improving task success and reducing token use~\cite{bhattaram2025geoflow}.
PowerChain~\cite{BADMUS2027113555} conditions distribution-grid analysis agents on expert-annotated, verified tool-call traces and achieves higher Pass@1 than unstructured retrieval baselines.
These results show that exemplar conditioning improves accuracy, but selecting the right exemplars for each query remains an open question.
Including all exemplars exceeds the context window, while including incorrect exemplars misleads the agent.

\subsection{Retrieval of Relevant Exemplars}

\noindent Retrieval-Augmented Generation (RAG) reduces prompt context by fetching task-relevant content via embedding similarity and conditioning the model on the retrieved results~\cite{lewis2020retrieval}.
Many systems implement dense retrieval by indexing exemplar candidates in a vector store and selecting the top-$k$ nearest neighbors under cosine similarity~\cite{Luo2023DrICLDI}.
Some deployments filter exemplar candidates using a minimum similarity threshold, either alone or combined with top-$k$, to exclude low-relevance exemplars~\cite{huang2024survey}.
For procedural exemplars that encode executable multi-step tool use, including call ordering, argument patterns, and state handoffs, fixed top-$k$ retrieval either omits relevant workflows for small $k$ or injects irrelevant context noise for large $k$~\cite{liu2024lost}.
For tool-calling and execution agents, retrieved content must encode executable structure, including call ordering, argument patterns, and prerequisite dependencies, rather than narrative descriptions~\cite{BADMUS2027113555}.

\subsection{Runtime Supervision of Tool Execution}
\noindent A separate challenge arises when tools execute in a stale or uninitialized state without raising exceptions, resulting in silent failures.
% These are primarily due to violations of tool dependencies and to erroneous environment states from tool mis-execution.
Dependency violations occur when the agent calls a state-dependent tool before its prerequisites have run, causing the tool to execute silently in the wrong state.
%In these scenarios, the agent receives spurious positive feedback, incorrectly believing that an erroneous action was successful while failing to recognize that prerequisite steps were bypassed due to insufficient environmental observability and system-state checks.
Prompting techniques can embed dependency constraints or negative examples in the prompt to steer tool selection~\cite{wu2024avatar}, but LLM sampling is probabilistic and cannot guarantee that all dependencies are satisfied.
Anthropic, the developer of Claude, similarly uses an LLM-advisor mechanism in its Claude Code agent harness, where the executor decides when to consult a stronger model for strategic guidance~\cite{anthropic2026advisor}.
Although this can improve reasoning, it remains stochastic and does not deterministically block prerequisite-invalid actions before execution.
Post-hoc self-correction methods revise actions in response to detected failures~\cite{shinn2023reflexion}, but cannot handle silent errors where a tool executes without raising exceptions on a stale state.

Several systems insert deterministic checks between the orchestrator and executor to verify preconditions before each call and return targeted feedback on violations.
ToolGate~\cite{liu2026toolgate} represents each tool as a state-transition rule with explicit entry requirements and expected state updates, then uses symbolic checks to verify that each call is admissible.
However, this precondition-based approach is difficult to apply when tools are external MCP services or closed solver APIs whose internals are inaccessible.
Pro2Guard~\cite{wang2025pro2guard} simulates candidate actions to predict violations before they occur.
AgentSpec~\cite{wang2025agentspec} defines event-triggered safeguard policies that require user confirmation before high-stakes actions (e.g., transfers to unverified recipients).
However, none of these systems track cross-tool state dependencies across a long-horizon execution.
They block unsafe individual actions but do not enforce multi-step workflow prerequisites.
For example, in power-system workflows, a tool can return a valid result while operating on the wrong feeder or solver state, without raising an exception.
Downstream tools can then silently read stale objects.
A prerequisite-enforcement mechanism must therefore operate at the tool interface level without requiring access to the tool's internals or source code.

\section{Preliminaries}
\label{sec:preliminaries}

%To develop a solution to the problem, we first need to discuss some preliminaries.

%\subsection{Agent}
%\label{sec:agent}
%\noindent An agent is a decision maker that selects and orchestrates tool calls to generate an executable workflow that produces the correct output for the query.
%Formally, after step $k$ produces $(a_k,o_k)$, the agent chooses the next action based on the interaction history $H_k=\{(a_1,o_1),\ldots,(a_k,o_k)\}$, i.e., $a_{k+1}=\pi_\theta(H_k)$ \cite{russell1995modern}.
%In our setting, each action $a_k$ specifies a tool name and bound arguments, i.e., $a_k=(n_k,\phi_k)$, where $n_k$ is the selected tool name and $\phi_k$ is an argument binding for that tool.
%When a large language model implements $\pi_\theta$, we refer to the agent as an LLM agent \cite{yao2023react}.

\subsection{Agentic AI System}
\label{sec:agentic_system}
\noindent An \textit{agentic AI system} couples an LLM-orchestrated agent with a stateful environment in a closed loop.
% An agent includes an \textit{orchestrator} to select tool call sequences, termed workflows, and an \textit{executor} to execute the workflow that produces an output to the query.
We model the agent as an \textit{orchestrator} that maps the interaction history and relevant context to the next tool-call action, and the environment as an \textit{executor} that executes the call and returns an observation.
%When a large language model implements $\pi_\theta$, we refer to the agent as an LLM agent \cite{yao2023react}.
At each step $k$, the agent proposes a tool-call action $a_k$.
The environment executes it in state $s_k$ to produce the next state $s_{k+1}$ and observation $o_k$.

\begin{center}
\begin{tabular}{c}
\texttt{Agent} $\xrightarrow{a_k}$ \texttt{Environment} $\xrightarrow{o_k}$ \texttt{Agent} $\xrightarrow{a_{k+1}} \cdots$
\end{tabular}
\end{center}

% \noindent Tools behave as deterministic functions of their arguments and the current environment state, but tool execution can modify $s_k$ through side effects.
% The resulting action--observation trace defines the executed workflow for the query.

\subsection{State Transition and Tool Classification}
\label{sec:tool-class}

\noindent Let $\mathcal{T}$ denote the set of available tools, and let $\mathcal{S}$ denote the environment-state space. 
At step $k$, the environment state is $s_k \in \mathcal{S}$.
Each tool $t\in \mathcal{T}$ defines a state-observation map $F_t:\mathcal{S}\times\Phi_t\to\mathcal{S}\times\mathcal{O}$.
Here, $\Phi_t$ is the argument space of tool $t$, and $\mathcal{O}$ is the observation space returned to the agent.
We classify tools by whether they modify the environment state: (i) A \textit{Write} tool updates persistent objects stored in the registry or simulator, meaning $s_{k+1}\neq s_k$ (e.g., updating the network state with solved voltage and current values). (ii) A \textit{Read} tool queries the current objects and returns an observation, leaving the state unchanged, meaning $s_{k+1}=s_k$ (e.g., reading node voltages). We label a tool as \textit{Write} if it mutates any persistent environment object and as \textit{Read} otherwise, and we validate this classification by comparing the environment state before and after each tool call. The before-and-after state test also catches latent synchronous side effects and labels such tools \textit{Write}. A \textit{Read} tool can execute correctly only after the required \textit{Write} tools have set up the state it reads. The workflow DAG (defined next) encodes this execution order.

\subsection{Workflow as a Directed Acyclic Graph (DAG)}
%\noindent The agent responds to query $q_u$ by executing tool invocations that satisfy state-dependent prerequisites and by using observations to guide later choices.
%We define the workflow as $w=\big(t_1(\phi_1),\ldots,t_m(\phi_m)\big)$, where $t_i\in T$ denotes an executable tool $i$ and $\phi_i$ denotes its arguments.
%We represent $w$ as a DAG $\mathcal{G}_w=(V_w,E_w)$, where vertices represents the tool calls with arguments $V_w=\{t_1(\phi_1),\ldots,t_m(\phi_m)\}$ and each directed edge $\big(t_i(\phi_i),t_j(\phi_j)\big)\in E_w$ encodes a prerequisite: $t_i(\phi_i)$ must execute before $t_j(\phi_j)$ for correctness.
%Any topological order of $\mathcal{G}_w$ yields an execution order provided $E_w$ captures all required prerequisites.
%For example, two invocations with the same parent and no dependency path between them can execute in either order (e.g., $t_4(\phi_4)$ and $t_5(\phi_5)$ in Fig.~\ref{fig:workflow_dag}), while an invocation can execute only after all of its incoming dependencies complete (e.g., $t_6(\phi_6)$ depends on both $t_4(\phi_4)$ and $t_5(\phi_5)$).
%Acyclicity rules out circular prerequisites and guarantees the existence of a topological order.

\noindent We define the \textit{workflow} for query $q$ as a sequence of tool invocations that satisfies all state-dependent prerequisites.
We formalize it as a directed acyclic graph (DAG), $\mathcal{G}_w = (V_w, E_w)$.
The vertex set $V_w = \{v_1, \dots, v_m\}$ represents distinct tool invocations, where each $v_i = t_i(\boldsymbol{\phi}_i)$ consists of a tool $t_i \in \mathcal{T}$ and its arguments $\boldsymbol{\phi}_i$. 
The edge set $E_w \subset V_w \times V_w$ encodes prerequisites.
A directed edge $(v_i, v_j) \in E_w$ requires $v_i$ to complete before $v_j$ executes.
For example, two invocations with the same parent and no dependency path between them can execute in either order (e.g., $t_4(\boldsymbol{\phi}_4)$ and $t_5(\boldsymbol{\phi}_5)$ in Fig.~\ref{fig:workflow_dag}).
Conversely, a tool executes only after all its predecessors complete (e.g., $t_6(\boldsymbol{\phi}_6)$ requires both $t_4(\boldsymbol{\phi}_4)$ and $t_5(\boldsymbol{\phi}_5)$).
Acyclicity rules out circular prerequisites and guarantees the existence of a topological order.
Acyclicity applies to individual call events, not to tool types.
Repeated calls to the same tool appear as distinct nodes, e.g., $t_p^{(1)}(\boldsymbol{\phi}_p^{(1)})\!\to t_q^{(1)}(\boldsymbol{\phi}_q^{(1)})\!\to t_p^{(2)}(\boldsymbol{\phi}_p^{(2)})$, which remains acyclic.
A cycle arises only if the same call event depends on itself, which a well-formed workflow excludes.

\begin{figure}[htbp]
\centering
\boldmath

\resizebox{0.53\columnwidth}{!}{
\begin{tikzpicture}[
    nd/.style={
        circle,
        draw=black,
        fill=white,
        drop shadow={
            shadow xshift=0.5mm,
            shadow yshift=-0.5mm
        },
        inner sep=4pt,
        font=\bfseries,
        line width=1.1pt
    },
    arrow/.style={
        -{Stealth[length=2.2mm,width=1.3mm]},
        line width=1.1pt,
        draw=black
    }
]

\node[nd] (n1) at (-3.5,0.0) {$t_1(\boldsymbol{\phi}_1)$};
\node[nd] (n2) at (-3.5,1.8) {$t_2(\boldsymbol{\phi}_2)$};
\node[nd] (n3) at (-1.8,0.9) {$t_3(\boldsymbol{\phi}_3)$};
\node[nd] (n4) at (0.2,1.8)  {$t_4(\boldsymbol{\phi}_4)$};
\node[nd] (n5) at (0.2,0.0)  {$t_5(\boldsymbol{\phi}_5)$};
\node[nd] (n6) at (2.2,0.9)  {$t_6(\boldsymbol{\phi}_6)$};
\node[nd] (n7) at (4.2,0.9)  {$t_7(\boldsymbol{\phi}_7)$};

\draw[arrow] (n1) -- (n3);
\draw[arrow] (n2) -- (n3);
\draw[arrow] (n3) -- (n4);
\draw[arrow] (n3) -- (n5);
\draw[arrow] (n4) -- (n6);
\draw[arrow] (n5) -- (n6);
\draw[arrow] (n6) -- (n7);

\end{tikzpicture}
}

\smallskip
{\footnotesize\bfseries
\centering
node $t_i(\boldsymbol{\phi}_i)\in V_w$: tool invocation\\
edge $(t_i(\boldsymbol{\phi}_i),t_j(\boldsymbol{\phi}_j))\in E_w$: ordering-based dependencies
\par}

\caption{Workflow as a directed acyclic graph (DAG). Nodes denote tool invocations, and directed edges encode ordering-based dependencies.}
\label{fig:workflow_dag}
\end{figure}

\subsection{\AWF{} (\texorpdfstring{$\mathcal{W}_\text{av}$}{W\_av})}
\label{sec:awf}
\noindent Power-systems experts solve distribution-grid analysis queries by executing validated tool-call sequences.
We encode these sequences as \awf{}.
Let $\mathcal{W}_\text{av}$ denote the available archive of these records.
Each record is a query-workflow record $(q^{(i)}, w^{(i)})$, where $q^{(i)}$ is a natural-language query and $w^{(i)}$
%=\big(t^{(i)}_1(\phi^{(i)}_1),\ldots,t^{(i)}_{m_i}(\phi^{(i)}_{m_i})\big)$
is the verified tool-call sequence that produces the correct output for $q^{(i)}$.
%We verify each record by running $w^{(i)}$ end-to-end in the same environment.

\section{An Agentic AI System: \PC{}}
\label{sec:PC_design}
%\noindent We now present the methodology used to solve the informal problem in Section \ref{sec:preliminaries}.
\newcommand{\Valid}{\mathcal{V}}

% \noindent - \textit{\textbf{Given}} a natural-language distribution grid analysis query $q_u$, a tool set $\mathcal{T}$ with schema set $\Sigma$ and associated data pipelines, a stateful execution environment, and \awf{} 
\noindent \textbf{Informal Problem:}

\noindent - \textit{\textbf{Given}} an \textit{unseen} distribution-grid analysis query $q_u$ in natural language, a tool set $\mathcal{T}$ and associated data pipelines, a stateful environment, and \awfs{} $\mathcal{W}_\text{av}$,

% \textbf{Goal} is to generate an executable computation graph (workflow): a partially ordered collection of tool calls with bound arguments that, when executed in the environment, produces the requested analysis outputs like an expert.

\noindent - \textbf{\textit{Goal}} is to return accurate analysis outputs for the query together with an executable workflow $w$ that generates them.
%The workflow serves as a verifiable certificate, as executing it in the environment reproduces the outputs.
\vspace{1mm}

\begin{figure}[htpb]
  \centering
  \includegraphics[width=0.76\columnwidth]{figures/supervisor.pdf}
    \caption{\PC{} execution architecture. The schema extractor summarizes the tool set $\mathcal{T}$, the retriever selects $\mathcal{W}_\text{sub}$ for the unseen query $q_u$, and the agent iterates with the JIT supervisor, which either blocks invalid actions with advisories or allows execution. The loop ends when the agent returns a final response.}
    \label{fig:supervisor}
    \end{figure}

\noindent \PC{} solves this problem as an agentic AI system for distribution-grid analysis built around two components: (i) an \textit{adaptive retriever} that selects query-relevant exemplars from $\mathcal{W}_\text{av}$ dynamically, and (ii) a \textit{Just-in-Time (JIT) supervisor} that enforces prerequisite constraints before each tool call.

\subsection{\PC{} System Architecture}
\label{sec:pc_arch}
\noindent Let $\Sigma$ denote the tool-schema set derived from the tool set $\mathcal{T}$.
%For each tool $t\in\mathcal{T}$, we define a schema $\sigma_t=\langle n_t,d_t,\Pi_t\rangle$, where $n_t$ is the tool name, $d_t$ is a short description, and $\Pi_t$ is the argument signature, which includes argument slots and their data-types.  $\boldsymbol{\phi}$ denotes the concrete argument values in a call.
For each tool $t\in\mathcal{T}$, we define a schema $\sigma_t=\langle n_t,d_t,\Pi_t\rangle$.
Here $n_t$ is the tool name, $d_t$ is a short description, and $\Pi_t$ is the argument signature, i.e., the list of input arguments and their data types.
An invocation of tool $t$ uses concrete values $\boldsymbol{\phi}_t$ for these parameters.
\begin{equation}
\Sigma := \{\sigma_t\}_{t\in\mathcal{T}}
\label{eq:toolkit}
\end{equation}
% We also maintain an archive $(\mathcal{W}_\text{av})$ of expert-generated \awf{} records.
% For a query $q_u$, an adaptive retriever selects a subset $\mathcal{W}_\text{sub}\subset\mathcal{W}_\text{av}$ and appends $\mathcal{W}_\text{sub}$ to the prompt alongside $q_u$ , interaction history $H$, and $\Sigma$.
The system also draws on a pre-built expert archive $\mathcal{W}_\text{av}$ of \awf{}.
For a query $q_u$, the adaptive retriever selects a subset $\mathcal{W}_\text{sub}\subset\mathcal{W}_\text{av}$.
The system constructs a prompt containing $q_u$, $\mathcal{W}_\text{sub}$, the interaction history $H_k$, and $\Sigma$.
Here, $H_k$ is the action-observation history up to step $k$.

The system then runs the closed-loop interaction between the agent and the environment, as shown in Fig.~\ref{fig:supervisor}.
Let $\pi_\theta$ denote the LLM policy, parameterized by $\theta$, that maps the current prompt context to a distribution over possible next actions.
At step $k$, the agent samples an action from this policy.
\begin{equation}
a_k \sim \pi_\theta\!\left(\cdot \,\middle|\, q_u,\mathcal{W}_\text{sub},H_k,\Sigma\right)
\label{eq:pc_action_sample}
\end{equation}
\noindent The agent either terminates and returns a final text response, or it returns a tool-call action $a_k=\langle n_k,\boldsymbol{\phi}_k\rangle$.
Here $n_k$ is the proposed tool name and $\boldsymbol{\phi}_k$ is its bound argument vector.
The resolver $\rho$ maps tool names to executable tools, so $t_k=\rho(n_k)$, and $\mathcal{E}$ denotes the environment executor.

In \PC{}, we place a JIT supervisor between the agent and the environment (see Fig.~\ref{fig:supervisor}).
The JIT supervisor checks the proposed tool-call action against the prerequisite rule library $\mathcal{C}$ before execution.
The supervisor is external to both the tools and the environment because the analysis tools may be connected as external MCP services whose source code cannot be modified.
Their state dependencies are nonetheless observable at the tool interface, so \PC{} runs prerequisite checks in the orchestration layer before any environment mutation occurs.
If the proposed call violates a rule, the supervisor blocks it and returns the advisory observation $o_k=\alpha(n_k)$ to the agent.
The supervisor does not execute the tool, and the environment state remains unchanged.
Because the check runs before execution, a blocked call produces no partial side effects, so the environment needs no transactional rollback.
Otherwise, the executor executes $t_k(\boldsymbol{\phi}_k)$ in the environment to produce a new state and observation $(s_{k+1},o_k)$, after which the system updates the history as $H_{k+1}=H_k\oplus(a_k,o_k)$.
%The executed workflow is the ordered list of tool calls that actually ran (with their bound arguments); for analysis, we also represent it as a DAG whose nodes are executed calls and whose directed edges encode prerequisite dependencies.
Sections~\ref{sec:smart-sel} and \ref{sec:jit-sup} detail the adaptive retriever and the JIT supervisor.

\subsection{Adaptive Retrieval of \AWFS{}}
\label{sec:smart-sel}

% \noindent Our goal is to develop an adaptive method to select \awfs{}   $\mathcal{W}_\text{sub}$ from the expert archive $\mathcal{W}_\text{av}$ that overcomes the drawbacks of works that choose top-$k$ exemplars, where $k$ is a fixed hyper-parameter\cite{badmus2025powerchain}.
% To that end, we develop (i) an adaptive cutoff computed from the ranked similarity profile and
% (ii) a second-stage filter that retains only exemplars whose workflows most mimic the tool execution parameters for $q_u$.
% (ii) a second-stage filter that keeps only exemplars whose workflows are structurally consistent with $q_u$,
% {\color{blue}
\noindent The adaptive retriever selects $\mathcal{W}_\text{sub}$ from $\mathcal{W}_\text{av}$ in two stages:
(i) Stage 1 applies an adaptive cutoff to the ranked similarity profile to produce a candidate set $\mathcal{W}_\text{cand}$, and (ii) Stage 2 filters $\mathcal{W}_\text{cand}$ by procedural relevance to $q_u$ to produce the final subset $\mathcal{W}_\text{sub}$.
% }
% The expert archive is $\mathcal{W}_\text{av}=\{(q_i,w_i)\}_{i=1}^N$, where each record pairs a stored query $q_i$ with its expert annotated workflow $w_i$.
% The context selector forms a text-similarity ranking in three steps: it encodes the unseen query $q_u$ and each stored query $q_i$ into embeddings $\psi(.)$, and computes cosine similarities between embeddings.
The expert archive is $\mathcal{W}_\text{av}=\{(q_i,w_i)\}_{i=1}^N$, where each record pairs a stored query $q_i$ with its expert-annotated workflow $w_i$.
 Here $N=|\mathcal{W}_\text{av}|$ is the number of archived exemplars.

Stage 1 uses only query-to-query matching.
The retriever excludes workflow traces because tool names, argument strings, and execution artifacts distort semantic similarity.
The adaptive retriever encodes $q_u$ and each stored query $q_i$ using the text encoder $\psi$, and computes cosine similarities. We re-index the archive such that the similarities are sorted in descending order:
\begin{align}
\mathrm{Sim}_i &= \cos(\psi(q_u),\psi(q_i)), \quad i=1,\ldots,N \label{eq:sim_i}\\
\mathrm{Sim}_{1} &\ge \mathrm{Sim}_{2} \ge \cdots \ge \mathrm{Sim}_{N} \label{eq:sim_rank}
\end{align}
where index $i$ denotes the exemplar's rank in descending order of similarity.

% The context selector treats $\{\mathrm{Sim}_{i}\}_{i=1}^N$ as a discrete relevance profile in descending order (i.e., most relevant exemplar appears first and the least appears last) and selects an adaptive cutoff (given by index $i_\text{cutoff}$ in ordered set) by fitting a two-segment least-squares model to the sequence and choosing the breakpoint that minimizes the weighted fit error:
The adaptive retriever finds cutoff $i_\text{cutoff}$ by fitting a two-segment least-squares line to the ranked profile $\{\mathrm{Sim}_{i}\}_{i=1}^N$ and selecting the breakpoint that minimizes the weighted fit error.
\begin{equation}\label{eq:elbow}
i_\text{cutoff} =\arg\min_{2\le b\le N-2}\Big[\frac{b}{N}\Phi(1,b)+\frac{N-b}{N}\Phi(b+1,N)\Big]
\end{equation}
Here $\Phi(r,s)$ denotes the root-mean-square error (RMSE) of the least-squares affine fit of $(i,\mathrm{Sim}_i)$ over $i=r,\ldots,s$.
Appendix~\ref{appx:cutoff} gives the full derivation of this objective.
% {\color{blue}
% while both the scorer and cutoff policy are selected empirically, Appendix \ref{appx:stage_selection} reports the full comparison across different.
% }
This yields the candidate set.
\begin{equation}\label{eq:cand}
\mathcal{W}_\text{cand}=\{(q_{i},w_{i})\}_{i=1}^{i_\text{cutoff}}
\end{equation}

Text-similarity ranking alone is insufficient because it can retrieve candidates whose query text resembles $q_u$ while their associated workflows do not match the tool structure implied by $q_u$.
% Conditioning on such mismatched exemplars can mislead the agent and reduce correctness.
% At retrieval time, the correct workflow for $q_u$ is unknown, so we cannot directly verify procedural alignment.
% We therefore apply an LLM-based filter that scores each candidate using both its query text and its expert-verified workflow trace, and retains only those that are procedurally consistent with $q_u$.
% The LLM acts only as a gate over a small candidate set; it selects among existing expert traces and does not synthesize a new tool sequence.
% This bounded discrimination task is simpler than long-horizon workflow generation.
Conditioning on such mismatched exemplars misleads the reasoning agent and degrades tool selection.
At retrieval time, the correct workflow for $q_u$ is unknown, so the retriever cannot verify procedural alignment directly.
We therefore apply an LLM-based filter that evaluates each candidate in $\mathcal{W}_\text{cand}$ using its query text and expert-verified workflow trace, retaining only those whose tool sequences match the requirements of $q_u$.
The LLM acts only as a gate over a small candidate set.
It selects among existing expert traces and does not synthesize a new tool sequence.
Selecting from a small fixed candidate set is simpler than generating a new tool sequence from scratch.
% \del{The two-stage design ensures the LLM filter in Stage~2 operates over a small candidate set rather than the full archive, which could exceed the context window.}
The two-stage design keeps the Stage 2 filter on a small candidate set rather than the full archive. Even when the archive fits the context window, conditioning on many irrelevant exemplars degrades selection accuracy and adds token cost and latency~\cite{liu2024lost}.

% The filter runs once at initialization and returns the subset $\mathcal{W}_\text{sub}$.
% \begin{equation}
% J = \Omega\big(\pi_{\theta}^{\mathrm{filter}}(\cdot \mid q_u, \mathcal{W}_\text{cand})\big), \quad
% \mathcal{W}_\text{sub} := \{(q_{(j)}, w_{(j)}) : j \in J\}
% \label{eq:filter_wsub}
% \end{equation}
% where $\Omega$ is a deterministic parser that converts the LLM output into an index set $J$.
% Appendix~\ref{app:filter_prompt} lists the exact filtering prompt.
% Fig.~\ref{fig:selector} illustrates the two-stage selection used to construct $\mathcal{W}_\text{sub}$.

\begin{figure}[htpb]
  \centering
  \includegraphics[width=0.50\columnwidth]{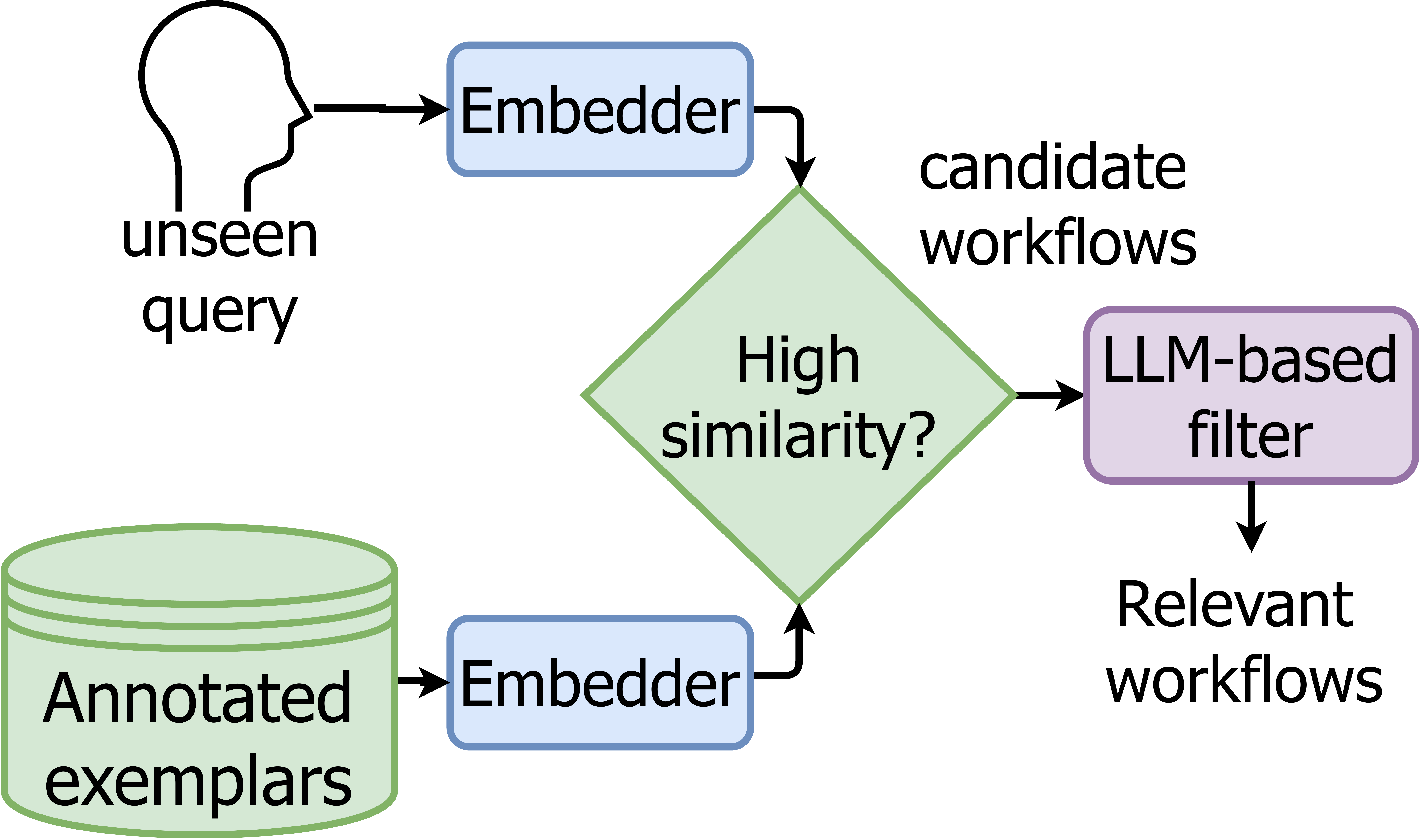}
    \caption{Two-stage exemplar selection. Stage 1 embeds the unseen query and archived queries from the annotated exemplars, ranks by cosine similarity, and applies a list-wise adaptive cutoff $i_\text{cutoff}$. Stage 2 then filters the candidate query-workflow records to produce $\mathcal{W}_\text{sub}$.}
    \label{fig:selector}
\end{figure}

The filter runs once at initialization and returns the subset $\mathcal{W}_\text{sub}$.
Here $\pi_{\theta}^{\mathrm{filter}}$ denotes the LLM policy for Stage 2 workflow filtering.
\begin{equation}
J = \Omega\big(\pi_{\theta}^{\mathrm{filter}}(\cdot \mid q_u, \mathcal{W}_\text{cand})\big), \quad
\mathcal{W}_\text{sub} := \{(q_{j}, w_{j}) : j \in J\}
\label{eq:filter_wsub}
\end{equation}
where $\Omega$ is a deterministic parser that converts the LLM output into an index set $J\subseteq\{1,\ldots,i_\text{cutoff}\}$, so the filter can only keep candidates from $\mathcal{W}_\text{cand}$.
Appendix~\ref{app:filter_prompt} lists the exact filtering prompt.
Fig.~\ref{fig:selector} illustrates the two-stage selection used to construct $\mathcal{W}_\text{sub}$.

% \begin{figure}[htpb]
%   \centering
%   \includegraphics[width=1\columnwidth]{figures/workflow_selector.pdf}
%     \caption{Two-stage exemplar selection. The system embeds $q_u$ and $\{q_i\}_{i=1}^N$, ranks by cosine similarity, selects an adaptive cutoff $i_\text{cutoff}$, and applies an LLM-based workflow filter to produce the final subset $\mathcal{W}_\text{sub}$.}
%     \label{fig:selector}
% \end{figure}

\subsection{Just-in-Time Supervision}
\label{sec:jit-sup}

% \noindent Even with relevant exemplars $\mathcal{W}_\text{sub}$, the agent can propose a tool call whose prerequisite \textit{Write} tools have not executed, causing it to act on stale state and corrupt downstream results. 
\noindent Even with relevant exemplars $\mathcal{W}_\text{sub}$, the agent has no direct view of the environment state.
It may therefore propose a tool call before the prerequisite tools have executed, corrupting downstream results.
We address this by inserting a JIT supervisor that checks each proposed tool call against a library of prerequisite rules $\mathcal{C}$ before any environment mutation occurs.
Because tool internals may be inaccessible (e.g., external MCP services or closed APIs), the supervisor operates at the tool interface level, where state dependencies remain observable.
We curate $\mathcal{C}$ from tool interface documentation and repeated precedence patterns in $\mathcal{W}_\text{av}$.

\textit{Trace-derived rule construction: }
To construct the rule library $\mathcal{C}$, the system automatically extracts candidate dependencies from repeated patterns in $\mathcal{W}_\text{av}$. A domain expert then verifies these candidates against the tool documentation. This build process occurs once per tool set, and the supervisor reuses the resulting library across all queries. For example, traces that call voltage-violation checks only after a power-flow solve imply that voltage checks require a solved network state.
These rules encode the most frequent prerequisite dependencies in $\mathcal{W}_\text{av}$, and the more rules in $\mathcal{C}$, the more invalid calls the supervisor can intercept.
The rules capture how experts already sequence the deployed tools, so $\mathcal{C}$ tracks the tool set in use rather than aiming to be exhaustive.
A tool without a rule remains executable and falls to the agent's exemplars and reasoning.
% Appendix \ref{appx:supervisor_diagnostics} reports rule coverage and advisory diagnostics .

Let $\mathcal{N}_{\mathrm{write}}$ denote the set of tool names for \textit{Write} tools.
At step $k$, let $\mathcal{N}^{\mathrm{exec}}_k$ denote the set of tool names already executed before the proposed action.
For each tool name $n$ covered by the supervisor, the rule specifies the required prerequisite \textit{Write} tools $\mathcal{N}^{\mathrm{req}}(n)\subseteq\mathcal{N}_{\mathrm{write}}$ and an advisory message $\alpha(n)$. 
\begin{equation}
c(n)=\langle n,\ \mathcal{N}^{\mathrm{req}}(n),\ \alpha(n)\rangle
\label{eq:constraint_tuple}
\end{equation}
We write $\mathrm{dom}(\mathcal{C})$ for the set of tool names that have a rule in $\mathcal{C}$. If a tool has no entry in $\mathcal{C}$ (i.e., $n\notin\mathrm{dom}(\mathcal{C})$), the supervisor does not block it.
% The evaluated rule library contains 112 rules governing 94 of the 108 tool interfaces, covering 1,097 of 1,330 tool-call events in the held-out expert workflows.
At step $k$, the agent proposes $a_k=\langle n_k,\boldsymbol{\phi}_k\rangle$.
The supervisor checks whether all prerequisites have executed, $\mathcal{N}^{\mathrm{req}}(n_k)\subseteq\mathcal{N}^{\mathrm{exec}}_k$.
It flags a violation when this fails.
\begin{equation}
\upsilon_k := \mathbf{1}\!\left[\,\mathcal{N}^{\mathrm{req}}(n_k)\nsubseteq\mathcal{N}^{\mathrm{exec}}_k\,\right]
\label{eq:violation_logic}
\end{equation}
Here $\mathbf{1}[\cdot]$ is the indicator function, and $\upsilon_k=1$ denotes a prerequisite violation at step $k$.
If $\upsilon_k=1$, the supervisor blocks the call and returns advisory $\alpha(n_k)$ without mutating the environment.
Otherwise, the call executes.
\begin{equation}
(s_{k+1},\, o_k) = \begin{cases} (s_k,\ \alpha(n_k)) & \text{if }\upsilon_k=1\\ 
\mathcal{E}(s_k,t_k(\boldsymbol{\phi}_k)) & \text{otherwise.} \end{cases}
\label{eq:intervention_logic}
\end{equation}
Here $\mathcal{E}$ denotes the environment executor, which executes the concrete tool call $t_k(\boldsymbol{\phi}_k)$ from state $s_k$ and returns the next state and observation.
Appendix~\ref{appx:supervisor_diagnostics} shows the advisory template.
Algorithm~\ref{alg:pc_execution} describes the end-to-end execution of \PC{}.
The operator $\oplus$ denotes append.

\begin{algorithm}[htpb]
\caption{\PC{} Algorithm}
\label{alg:pc_execution}
\resizebox{\linewidth}{!}{%
\begin{minipage}{1.4\linewidth}
\begin{algorithmic}[1]
% ADDED: Explicitly stating T contains the tools t
\Require query $q_u$, tool set $\mathcal{T}$ with its schemas $\Sigma$, \awfs{} $\mathcal{W}_\text{av}$, prerequisite rules $\mathcal{C}$, stateful environment.
\Ensure executed workflow trace $w$ and final response

\State \textbf{\textit{Phase 1: Adaptive Retrieval}}
    \State \hspace{0.5em} \textbf{Compute similarities} $\{\mathrm{Sim}_i\}_{i=1}^N$ and \textbf{rank and re-index}  \Comment{Eqs. \eqref{eq:sim_i},\eqref{eq:sim_rank}}
    \State \hspace{0.5em} \textbf{Calculate cutoff} $i_\text{cutoff}$, $\mathcal{W}_\text{cand} \gets \{(q_{i},w_{i})\}_{i=1}^{i_\text{cutoff}}$ \Comment{Eq. \eqref{eq:elbow},\eqref{eq:cand}} 
    \State \hspace{0.5em} \textbf{Filter} $J \gets \Omega(\pi_\theta^{\mathrm{filter}}(\cdot \mid q_u, \mathcal{W}_\text{cand}))$, $\mathcal{W}_\text{sub} \gets \{(q_{j},w_{j}) : j \in J\}$
    \Comment{Eq. \eqref{eq:filter_wsub}}

\State \textbf{\textit{Phase 2: Just-in-Time} supervision}
\State \textbf{Init} $s_0\gets\emptyset$, $H_0\gets()$, $w\gets()$, $\mathcal{N}^{\mathrm{exec}}_0\gets\emptyset$, $k\gets0$

\Loop
    \State \textbf{Action} $a_k \sim \pi_\theta(\cdot \mid q_u,\mathcal{W}_\text{sub},H_k,\Sigma)$ \Comment{Eq. \eqref{eq:pc_action_sample}}

    \If{$a_k$ is not final}
        \State \textbf{Extract} $\langle n_k,\boldsymbol{\phi}_k\rangle$ from $a_k$
        \If{$n_k \notin \mathrm{dom}(\mathcal{C})$} \Comment{no prerequisite rule}
            \State \textbf{Execute \& Observe} $(s_{k+1},o_k)\gets \mathcal{E}\!\left(s_k,\,t_k(\boldsymbol{\phi}_k)\right)$
            \State \textbf{Update} workflow $w \gets w \oplus a_k$
            \State $\mathcal{N}^{\mathrm{exec}}_{k+1} \gets \mathcal{N}^{\mathrm{exec}}_k\cup\{n_k\}$
        \Else
            \State \textbf{Set} $\upsilon_k \gets \mathbf{1}\!\left[\mathcal{N}^{\mathrm{req}}(n_k)\nsubseteq \mathcal{N}^{\mathrm{exec}}_k\right]$
            \Comment{Eq. \eqref{eq:violation_logic}}

            \If{$\upsilon_k=1$}
                \State \textbf{Observation} (advisory) $o_k \gets \alpha(n_k)$ \Comment{Eq. \eqref{eq:intervention_logic}}
                \State \textbf{Freeze} state: $s_{k+1} \gets s_k$ \Comment{no environment state change}
                \State $\mathcal{N}^{\mathrm{exec}}_{k+1} \gets \mathcal{N}^{\mathrm{exec}}_k$
            \Else
                \State \textbf{Execute \& Observe} $(s_{k+1},o_k)\gets \mathcal{E}\!\left(s_k,\,t_k(\boldsymbol{\phi}_k)\right)$
                \State \textbf{Update} workflow $w \gets w \oplus a_k$
                \State $\mathcal{N}^{\mathrm{exec}}_{k+1} \gets \mathcal{N}^{\mathrm{exec}}_k\cup\{n_k\}$
            \EndIf
        \EndIf

        \State \textbf{Append} history $H_{k+1}\gets H_k\oplus(a_k,o_k)$
        \State $k\gets k+1$
    \Else
        \State \Return $w$, final response
    \EndIf
\EndLoop

\end{algorithmic}
\end{minipage}%
}
\end{algorithm}

\textit{Analysis of workflow space under supervision:}
We analyze how supervision reduces the candidate workflow space when agents follow supervisor advisories.
Without bounds on tool selection or repetition, the number of possible workflow DAGs is unbounded.
To calculate a baseline count of possible workflows, we first assume that the agent calls each tool at most once and we ignore tool arguments.
Let $D_m$ denote the number of labeled acyclic digraphs on $m$ tool-call events~\cite{robinson2006counting}, where $m \le |\mathcal{T}|$ is the number of tools selected.
The total workflow DAG count over all possible tool subsets is
\begin{equation}
W(|\mathcal{T}|) =
\sum_{m=0}^{|\mathcal{T}|}
\binom{|\mathcal{T}|}{m} D_m 
\label{eq:base_workflow_count}
\end{equation}
This count grows rapidly with $|\mathcal{T}|$.
This growth matters in practice because a stochastic LLM can propose many syntactically valid but prerequisite-invalid execution orders.
The supervisor prevents such execution orders from reaching the executor by allowing only DAGs that satisfy the prerequisite constraints in $\mathcal{C}$.
For any subset of tools $S \subseteq \mathcal{T}$ of size $m$, let $D^{\mathcal{C}}(S)$ denote the number of labeled acyclic digraphs on $S$ that satisfy all prerequisite rules in $\mathcal{C}$. By construction, $D^{\mathcal{C}}(S)\le D_m$.
The supervised workflow DAG count is
\begin{equation}
W^{\mathcal{C}}(|\mathcal{T}|) =
\sum_{m=0}^{|\mathcal{T}|}
\sum_{\substack{S \subseteq \mathcal{T} \\ |S| = m}} D^{\mathcal{C}}(S)
\le W(|\mathcal{T}|)
\label{eq:supervised_base_count}
\end{equation}

We next allow repeated tool calls and argument choices.
Let $R_i$ be the maximum number of calls to tool $t_i$, let $\mathbf{R}=(R_1,\ldots,R_{|\mathcal{T}|})$, let $\Valid_i$ be the admissible argument set for tool $t_i$, and let $L$ be the maximum workflow length.
For a length-$m$ workflow, let $c_i$ be the number of calls to tool $t_i$, so $\mathbf{c}=(c_1,\ldots,c_{|\mathcal{T}|})$ and $\sum_i c_i=m$.
The flag $\xi\in\{0,1\}$ controls whether repeated call placements are counted.
The unsupervised count is
\begin{equation}
W_{L,\mathbf{R}}^{\xi}
 =
\sum_{m=0}^{L}
D_m
\sum_{\substack{
c_1+\cdots+c_{|\mathcal{T}|}=m\\
0\le c_i\le R_i
}}
\left(
\frac{m!}{c_1!\cdots c_{|\mathcal{T}|}!}
\right)^{\xi}
\prod_{i=1}^{|\mathcal{T}|}|\Valid_i|^{c_i}
\label{eq:extended_count}
\end{equation}
When $\xi=1$, we count each assignment of the $c_i$ calls of each tool $t_i$ to the $m$ call-event positions as distinct.
When $\xi=0$, we count each tool-count vector $\mathbf{c}$ once, ignoring which positions hold the repeated calls.
The supervised count is
\begin{equation}
W_{L,\mathbf{R}}^{\mathcal{C},\xi}
 =
\sum_{m=0}^{L}
\sum_{\substack{
c_1+\cdots+c_{|\mathcal{T}|}=m\\
0\le c_i\le R_i
}}
\left(
\frac{m!}{c_1!\dots c_{|\mathcal{T}|}!}
\right)^{\xi}
D^{\mathcal{C}}(\mathbf{c})
\prod_{i=1}^{|\mathcal{T}|}|\Valid_i|^{c_i}
\label{eq:supervised_count}
\end{equation}
where $D^{\mathcal{C}}(\mathbf{c})$ is the number of labeled acyclic digraphs on the $m$ tool-call events specified by the count vector $\mathbf{c}$ that satisfy $\mathcal{C}$. Since $D^{\mathcal{C}}(\mathbf{c})\le D_m$ termwise for any $\mathbf{c}$, $W_{L,\mathbf{R}}^{\mathcal{C},\xi}\le W_{L,\mathbf{R}}^{\xi}$.
Each prerequisite rule in $\mathcal{C}$ weakly reduces the number of workflow DAGs that can reach the executor.
For example, with two tools $A$ and $B$, the unsupervised DAG space contains six workflows.
These are $\emptyset$, $A$, $B$, $A; B$, $A\to B$, and $B\to A$, where $A; B$ denotes a workflow containing both calls with no dependency edge between them.
If the supervisor enforces $A$ as a prerequisite for $B$, only $\emptyset$, $A$, and $A\to B$ remain valid, reducing the space from $6$ to $3$.

\section{Experimental Setup}
\label{sec:setup}
%{\color{red} \noindent This section defines the queries, baselines, models, and scoring protocol used to measure \PC{} correctness and efficiency.}

\noindent This section describes the benchmark queries, exemplar archive, agentic systems, LLMs, and evaluation metrics used to measure \PC{} correctness and efficiency.

\subsection{Queries and Study Scope}
\label{sec:query_scope}
%{\color{red}
%\noindent We evaluate \PC{} on 30 expert-curated distribution-grid queries (Appendix \ref{appx:queries}).
%The query set spans four analysis classes: (i) \textit{data-only query}, (ii) \textit{unbalanced three-phase power flow} \cite{pandey2018robust}, (iii) \textit{dynamic hosting capacity} with $\ell_1/\ell_2/\ell_{\infty}$ formulations \cite{badmus2024anoca,liu2022using,moring2023inexactness,yi2022fair}, and (iv) \textit{three-phase infeasibility analysis} with $\ell_1$- and $\ell_2$-objective variants \cite{foster2022three,panthee2025solving}.
%The queries vary in workflow length from $\le 3$ tool calls to $>10$ tool calls.
%%For each query, we define and verify a ground-truth workflow that serves as the scoring reference.
%%We next describe how we curate the \awf{} library used for query-conditioned in-context learning.
%}

% {\color{blue}

\noindent  
% \del{We evaluate \PC{} on real-world distribution feeders from Vermont Electric Cooperative (VEC).}
We evaluate \PC{} on four real-world distribution feeders in Vermont, United States (Rochester, Stowe, Glover, and South Hero), each provided as a GridLAB-D network model with bus coordinates and AMI load time series.
Using these feeders and associated grid-analysis data, we curate 200 expert-verified query-workflow records across ten distribution-grid task families.
These records span tasks ranging from reading network information to performing distribution-grid analyses, limit checks, generating plots, and exporting results.
Their workflows range from 2 to 21 tool calls.
We then reserve 50 records for the exemplar archive $\mathcal{W}_\text{av}$ and use the remaining 150 records as held-out evaluation queries.
Appendix~\ref{appx:queries} gives the task-family breakdown and representative queries.
We split the data within-family.
The exemplar archive and evaluation set cover the same task families, but no evaluation query appears in the archive $\mathcal{W}_\text{av}$. 
The held-out queries also differ from the exemplar records in feeder choice, component parameters, time windows, solver settings, physical limits, and required artifacts. 
We therefore evaluate how well an agent adapts expert distribution-grid procedures within the studied families.
% }

\subsection{Curation of \AWF{}}
\label{sec:task_workflow_curation}
%{\color{red}
%\noindent We curate 50 \awf{} for in-context learning. 
%Each pair includes an expert-written query along with a fully verified workflow.
%All 50 \awf{} are distinct from the 30 evaluation queries in Section \ref{sec:query_scope}.
%10 of the 50 \awf{} cover the same four classes of analyses as the 30 unseen queries.
%To evaluate the retrieval accuracy of \PC{}, the remaining 40 \awf{} represent analyses different from four classes of analyses in 30 unseen queries.
%Code and data are available at 
%% \url{https://github.com/emmanuelbadmus/PowerDAG}
%\url{https://github.com/\PC/\PC}
%}

\noindent We build the exemplar archive $\mathcal{W}_\text{av}$ from the 50 reserved records, with 5 records per task family.
Each record pairs a natural-language query with the verified tool-call workflow that completes it in the evaluation environment.
We supply these records to the agent as in-context exemplars.
Appendix~\ref{appx:awf} shows the record format and representative records.
%To evaluate exemplar selection, we treat archive records from the same task family as relevant matches for an evaluation query. 
%This same-family comparison lets us measure whether the two-stage selection process in Section~\ref{sec:smart-sel} retrieves and filters task-relevant records from the archive.
Appendix~\ref{appx:stage_selection} reports the scorer comparison, cutoff-policy comparison against fixed top-$k$ retrieval, and Stage 2 workflow filtering results, including candidate counts and selected exemplars.

% \subsection{Execution Environment and Tool Catalog}
% \label{sec:analysis_tools}

% \noindent To ensure a controlled comparison, we keep the environment, data pipelines, and tool sets same across all agentic systems .
% The complete catalog consists of 82 tools; however, the 30 evaluation queries require only 21 specific tools (covering data loading, solver runs, constraint checks, plotting, and export). 
% We intentionally retain the remaining 61 tools during evaluation to test the agents' ability to select correct tools within an expanded action space.

% \subsection{Networks and Data Interfaces}
% \label{sec:benchmark_scope}
% \noindent We run all experiments on Vermont Electric Cooperative (VEC) distribution feeders.
% We provide feeder and time-series inputs with three callable tools in the catalog: \texttt{load\_network(args)} loads topology and network objects from GeoJSON and GridLAB-D files for a given feeder; \texttt{fetch\_ami(args)} retrieves meter-level AMI load time series for a specified window for specified meter ids; \texttt{load\_solar(args)} constructs PV generation profiles from timestamp, location, irradiance, and nameplate capacity.
% %These tools fix topology, load, and PV inputs across all systems.

\subsection{Agentic Systems, Tools, and Evaluation Setup}
\label{sec:analysis_tools}
%{\color{red}
%\noindent Table~\ref{tab:baselines} summarizes the systems compared that we use in our evaluation, along with the available features.
%To ensure a controlled comparison, we keep the environment, data pipelines, and tool sets fixed across all agentic systems.
%The full tool set contains 82 tools; the 30 evaluation queries require only 21 tools spanning data loading, solver runs, constraint checks, plotting, and export.
%We intentionally added 61 tools during evaluation to test whether agents select the correct tools within an expanded action space.
%% We run all experiments on Vermont Electric Cooperative (VEC) distribution feeders.
%We run all experiments on real-world distribution feeders (we do not name in this version due to double-blind review).
%Data pipelines include the AMI database, GeoJSON GIS files, time-series PV generation data, and network data in GridLAB-D format corresponding to the same real-world utility.
%%Feeder topology and time-series inputs enter the environment through three callable tools: \texttt{load\_network(args)} loads feeder topology and network objects from GeoJSON and GridLAB-D files; \texttt{fetch\_ami(args)} retrieves meter-level AMI load time series over a specified window for specified meter IDs; and \texttt{load\_solar(args)} constructs PV generation profiles from timestamp, location, irradiance, and nameplate capacity.
%}

% {\color{blue}
\noindent Table~\ref{tab:baselines} summarizes the six agentic configurations evaluated in this study.
They include three no-exemplar, no-supervision baselines (ReAct, LangChain, CrewAI), two single-component variants, the \textit{PowerChain-AR agent} (adaptive retrieval without JIT supervision) and the \textit{JIT-supervised agent} (JIT supervision without exemplars), and the full \PC{} system.
We use the same environment, data pipelines, and  tool set across all systems, so performance differences reflect agent design rather than differences in tools or data.
We expose 108 tools to the agent spanning feeder loading, network analysis, optimization, plotting, and export, sufficient to cover all ten task families in Section~\ref{sec:query_scope}.
For each query, the agent must select the needed tools with argument bindings from this full tool set and call them in the required order.
%The simulator sits behind the tools. %Supporting another simulator means adding its loader and analysis tools, with retrieval and supervision unchanged and the prerequisite rules re-curated per tool set.
Appendix~\ref{appx:supervisor_diagnostics} reports which tools are checked by the JIT supervisor and how often advisories are triggered.
%We start each model--method--query run with a fresh agent instance, empty interaction history, and a reset execution environment.
%For systems with JIT supervision, we also reset the supervisor live-fact set before the run begins. 
%We save generated artifacts under the corresponding query/run output path and do not carry failed-run state into later runs.
% }

\begin{table}[htpb]
\caption{Comparison of Agentic Systems and Enabled Components}
\label{tab:baselines}
\centering
\scriptsize
\setlength{\tabcolsep}{2pt}
\resizebox{\linewidth}{!}{
\begin{tabular}{lcccccc}
\toprule
\textbf{Component} & 
\shortstack{\textbf{ReAct}\\\textbf{Baseline}} & 
\shortstack{\textbf{LangChain}\\\textbf{ReAct}} & 
\shortstack{\textbf{CrewAI}\\\textbf{Single-Agent}} & 
\shortstack{\textbf{PowerChain-AR}\\\textbf{Agent}} & 
\shortstack{\textbf{JIT-supervised}\\\textbf{Agent}} & 
\shortstack{\textbf{\PC{}}\\\textbf{Agent}} \\
\midrule
Tool descriptions          & \cmark & \cmark & \cmark & \cmark & \cmark & \cmark \\
Action-observation loop    & \cmark & \cmark & \cmark & \cmark & \cmark & \cmark \\
Workflow exemplars         & \xmark & \xmark & \xmark & \cmark & \xmark & \cmark \\
Adaptive Stage 1 retrieval & \xmark & \xmark & \xmark & \cmark & \xmark & \cmark \\
Stage 2 workflow filtering & \xmark & \xmark & \xmark & \cmark & \xmark & \cmark \\
JIT supervisor             & \xmark & \xmark & \xmark & \xmark & \cmark & \cmark \\
\bottomrule
\end{tabular}
}
\end{table}

\subsection{LLMs and Compute Platforms}
\label{sec:models}
%{\color{red}
%\noindent We evaluate two proprietary LLMs, \llm{GPT-4o-mini} \cite{openai_gpt4omini_2026} and \llm{GPT-5.2} \cite{openai_gpt52_2026}, via the OpenAI API \cite{openai_models_2026}.
%We also evaluate four open-weight LLMs: \llm{Qwen3-14B} \cite{qwen3_14b_card_2025}, \llm{Qwen3-32B} \cite{qwen3_32b_card_2025}, \llm{gpt-oss-20b} \cite{openai_gptoss_card_2025}, and \llm{gpt-oss-120b} \cite{openai_gptoss_card_2025}.
%We serve the open-weight models with the \texttt{vLLM} server \cite{vllm_openai_server_2026} on a node with $2\times$ NVIDIA H100 SXM GPUs (80 GB each) \cite{nvidia_h100_2025}.
%For exemplar selection, we embed queries and archived exemplar texts using OpenAI text embedder \llm{text-embedding-3-large} \cite{openai_text_embedding_3_large_docs}.
%}

\noindent We evaluate the agentic systems in Table \ref{tab:baselines} across 10 large language models (LLMs), comprising 5 proprietary and 5 open-weight models.
% The proprietary models are \llm{GPT-4o Mini} \cite{openai_gpt4omini_2026}, \llm{GPT-5.5} \cite{openai_gpt55_2026}, \llm{Gemini 3.1 Flash-Lite} \cite{google_gemini31_flash_lite_card_2026}, \llm{Gemini 3.1 Pro} \cite{google_gemini31_pro_card_2026}, and \llm{Claude Haiku 4.5} \cite{anthropic_haiku45_system_card_2025}. 
% We access these models through the OpenAI, Gemini, and Claude APIs \cite{openai_models_2026,google_gemini_api_docs_2026,anthropic_models_overview_2026}. 
% The open-weight models are \llm{Llama 3.2-3B-Instruct} \cite{llama32_3b_instruct_card_2024}, \llm{Gemma-4 E2B} \cite{gemma4_e2b_card_2026}, \llm{Gemma-4 31B} \cite{gemma4_31b_card_2026}, \llm{GPT-OSS 120B} \cite{openai_gptoss_card_2025}, and \llm{Qwen3.6-27B} \cite{qwen36_27b_card_2026}. 
% We serve the open-weight models with an OpenAI-compatible \texttt{vLLM} endpoint \cite{vllm_openai_server_2026} on a node with $2\times$ NVIDIA H100 SXM GPUs, each with 80 GB of memory \cite{nvidia_h100_2025}.
The proprietary models are \llm{GPT-4o Mini}, \llm{GPT-5.5}, \llm{Gemini 3.1 Flash-Lite}, \llm{Gemini 3.1 Pro}, and \llm{Claude Haiku 4.5}.
We access these models through the OpenAI, Gemini, and Claude APIs.
The open-weight models are \llm{Llama 3.2-3B-Instruct}, \llm{Gemma-4 E2B}, \llm{Gemma-4 31B}, \llm{GPT-OSS 120B}, and \llm{Qwen3.6-27B}.
We serve the open-weight models with an OpenAI-compatible \texttt{vLLM} endpoint on a node with $2\times$ NVIDIA H100 SXM graphics processing units (GPUs), each with 80 gigabytes (GB) of memory.
Open-weight models run at temperature 0.0.
Most proprietary models, especially reasoning models such as \llm{GPT-5.5}, do not expose an adjustable temperature. 
We therefore use each proprietary model's API-default decoding.
We leave top-$p$, frequency penalties, presence penalties, and maximum generation length at provider defaults.
The Stage 2 exemplar filter runs at temperature 0.

For Stage 1 exemplar retrieval, we use Google \llm{gemini-embedding-2} \cite{google_gemini_embedding2_2026} with cosine similarity because it gives the best overall retrieval score in Appendix~\ref{appx:stage_selection}. 
For Stage 2 workflow filtering, we use the same LLM backend that evaluates the corresponding end-to-end run.
This setup separates fixed retrieval from model-specific filtering.
Stage 1 uses the same embedder for all models, so Stage 2 selects from an already relevant candidate pool.
Even smaller models benefit from these exemplars.
Gemma-4 E2B rises from $5.3\%$ without exemplars to $16.7\%$ with them.
 %Thus, each LLM backend runs Stage 2 exemplar filtering before it begins executing the task.
Table~\ref{tab:main_results} reports the tokens used during successful agent runs, and Appendix~\ref{appx:stage_selection} reports the retrieval and workflow filtering diagnostics.

\subsection{Benchmark Evaluation and Metrics}
\label{sec:metrics}

\noindent We evaluate all systems on the 150 held-out queries described in Section~\ref{sec:query_scope}.
We run each query with 10 LLMs and 6 agent-system configurations, yielding 9,000 runs.
In each run, the agent calls tools, observes tool outputs, and continues until it returns a final answer or reaches a stop condition.
We follow the $\tau$-bench evaluation protocol~\cite{yao2024taubench}, which scores each independent run separately rather than averaging across retries.
The evaluator scores each run against its corresponding executable expert workflow.
The evaluator checks tool coverage, argument compatibility, dependency order, and required validation or export calls.

We report two correctness metrics, Pass@1 and Precision (Pr).
Pass@1 measures task success.
A run passes if the executed workflow covers all required expert tool calls with compatible arguments and a valid dependency order.
Precision measures trace equivalence.
It checks whether the agent's tool-call sequence matches the expert sequence after DAG normalization.
Pass@1 tolerates extra calls to \textit{Read} tools but not extra \textit{Write} calls that change the environment state.
The only repeatable state-changing tools are limit setters, which the evaluator deduplicates to their final values.

\subsubsection{Pass@\texorpdfstring{$k$}{k} (P@\texorpdfstring{$k$}{k})}
Pass@$k$ measures the probability that at least one of $k$ generated workflows for a query satisfies the success criterion~\cite{chen2021evaluating}.
For a held-out query $q$, let $G_q$ denote the number of generated workflows and let $S_q$ denote the number of successful workflows.
The general Pass@$k$ estimator for that query is
\begin{equation}
\text{Pass@}k =
1 -
\frac{\binom{G_q-S_q}{k}}{\binom{G_q}{k}},
\qquad 1 \le k \le G_q
\label{eq:pass_at_k}
\end{equation}
In this study, for each reported model and method pair, we run each held-out query once, so we report the $k=1$ case over $\mathcal{Q}_{\mathrm{eval}}$:
\begin{equation}
\text{P@1} =
\frac{1}{|\mathcal{Q}_{\mathrm{eval}}|}
\sum_{q\in\mathcal{Q}_{\mathrm{eval}}}
\mathbb{I}\!\left[\text{query } q \text{ succeeds}\right]
\label{eq:pass_at_one}
\end{equation}
Here $|\mathcal{Q}_{\mathrm{eval}}|=150$, and $\mathbb{I}[\cdot]$ is the indicator function, equal to $1$ when the condition is true and $0$ otherwise.

\subsubsection{Precision (Pr)}
Precision measures whether the agent workflow matches the expert workflow after DAG normalization.
For each held-out query $q\in\mathcal{Q}_{\mathrm{eval}}$, let $\hat{w}_q$ denote the agent-executed workflow and let $w_{e,q}$ denote the corresponding expert workflow.
We compute Precision as
\begin{equation}
\text{Pr} =
\frac{1}{|\mathcal{Q}_{\mathrm{eval}}|}
\sum_{q\in\mathcal{Q}_{\mathrm{eval}}}
\mathbb{I}\!\left[\hat{w}_q \equiv w_{e,q}\right]
\label{eq:precision_metric}
\end{equation}
The evaluator fixes each dependency DAG order, so valid reorderings with nonconflicting read/write sets receive the same score.

\subsubsection{Token use (Tk)}
We report mean token use over successful runs.
For each run, token usage is the sum of prompt and completion tokens across all model calls during agent execution.
Table~\ref{tab:main_results} reports this mean in thousands of tokens.

% \textbf{Token use.}
% Efficiency measures total model token use, including both input prompt tokens and output completion tokens.
% For each run $i$, let $T_i$ denote the sum of prompt and completion tokens.
% We compute the mean total tokens over successful executions as
% \begin{equation}
% \text{Tk}=\frac{1}{|\mathcal{I}_{\mathrm{pass}}|}
% \sum_{i\in\mathcal{I}_{\mathrm{pass}}}T_i,
% \quad
% \mathcal{I}_{\mathrm{pass}}:=\{i:\text{run } i \text{ passes}\}.
% \end{equation}
% Table~\ref{tab:main_results} reports this value in thousands.

The evaluator implementation is available at \url{https://github.com/emmanuelbadmus/DistGrid-AgentBench}, and the \PC{} agent at \url{https://github.com/emmanuelbadmus/PowerDAG}.

\section{Results}
\label{sec:results}
%{\color{red}
%\noindent This section reports benchmark results for \PC{} on the unseen distribution-grid analysis queries. 
%Table \ref{tab:main_results} compares \PC{} against four baselines (CrewAI, LangChain, Vanilla ReAct, and PowerChain \cite{badmus2025powerchain}) across six LLMs. 
%%We measure correctness using Pass@1, Pass@3, and workflow Precision, and efficiency using Tokens-per-Pass@1.
%\PC{} achieves the highest Pass@1 and Pass@3 rates across all models. 
%It also uses fewer tokens per successful run. 
%The improvements are largest on smaller models (GPT-4o-mini, GPT-OSS-20B, Qwen3-14B), where other agents fail to complete multi-step workflows.
%We analyze these results in four parts: baseline comparisons, model scaling behavior, efficiency analysis, and ablation studies.
%}

\noindent We report Pass@1, Precision, and token use for \PC{} on the 150-query benchmark.
We report Pass@1 and Precision as percentages.
Table~\ref{tab:main_results} shows six agentic configurations across ten LLMs.
They include three no-exemplar, no-supervision baselines (ReAct, LangChain, CrewAI), two ablations that isolate the proposed mechanisms (the PowerChain-AR agent and the JIT-supervised agent), and the full \PC{} system.

\begin{table*}[!t]
\centering
\caption{Comparison of agentic systems across Pass@1, Precision (\%), and token use.}
\label{tab:main_results}
\footnotesize
\setlength{\tabcolsep}{1.6pt}
\renewcommand{\arraystretch}{1.05}
\newcommand{\modelhead}[3]{\multicolumn{3}{c#1}{\shortstack[c]{\llm{#2}\\\llm{#3}}}}
\begin{minipage}{\textwidth}
\centering
\resizebox{\linewidth}{!}{%
\begin{tabular}{@{}l*{5}{ccc}|*{5}{ccc}@{}}
\toprule
& \multicolumn{15}{c|}{\textbf{Open-Weight Models}}
& \multicolumn{15}{c}{\textbf{Proprietary Models}} \\
\cmidrule(lr){2-16}\cmidrule(lr){17-31}

& \modelhead{}{Llama}{3.2-3B}
& \modelhead{}{Gemma-4}{E2B}
& \modelhead{}{GPT-OSS}{120B}
& \modelhead{}{Gemma-4}{31B}
& \modelhead{|}{Qwen3.6}{27B}
& \modelhead{}{GPT-4o}{Mini}
& \modelhead{}{Gemini 3.1}{Flash-Lite}
& \modelhead{}{Claude}{Haiku 4.5}
& \modelhead{}{Gemini 3.1}{Pro}
& \modelhead{}{GPT}{5.5} \\

\cmidrule(lr){2-4}
\cmidrule(lr){5-7}
\cmidrule(lr){8-10}
\cmidrule(lr){11-13}
\cmidrule(lr){14-16}
\cmidrule(lr){17-19}
\cmidrule(lr){20-22}
\cmidrule(lr){23-25}
\cmidrule(lr){26-28}
\cmidrule(lr){29-31}

\textbf{Method}
& \textbf{P@1} & \textbf{Pr} & \textbf{Tk}
& \textbf{P@1} & \textbf{Pr} & \textbf{Tk}
& \textbf{P@1} & \textbf{Pr} & \textbf{Tk}
& \textbf{P@1} & \textbf{Pr} & \textbf{Tk}
& \textbf{P@1} & \textbf{Pr} & \textbf{Tk}
& \textbf{P@1} & \textbf{Pr} & \textbf{Tk}
& \textbf{P@1} & \textbf{Pr} & \textbf{Tk}
& \textbf{P@1} & \textbf{Pr} & \textbf{Tk}
& \textbf{P@1} & \textbf{Pr} & \textbf{Tk}
& \textbf{P@1} & \textbf{Pr} & \textbf{Tk} \\
\midrule

ReAct Baseline
& 6.67 & 0.00 & 128
& 2.00 & 2.00 & 23
& 41.33 & 21.33 & 62
& 47.33 & 1.33 & 44
& 48.67 & 18.67 & 112
& 20.67 & 1.33 & 45
& 40.67 & 14.67 & 69
& 36.00 & 13.33 & 69
& 88.00 & 42.67 & 104
& 77.33 & 36.00 & 84 \\

LangChain ReAct
& 4.00 & 0.00 & 134
& 1.33 & 1.33 & 47
& 36.00 & 18.67 & 63
& 38.00 & 0.00 & 42
& 43.33 & 16.00 & 114
& 25.33 & 4.67 & 74
& 36.00 & 13.33 & 57
& 43.33 & 22.67 & 69
& 80.00 & 31.33 & 89
& 74.67 & 48.00 & 93 \\

CrewAI Single-Agent
& 2.00 & 0.00 & 100
& 0.67 & 0.67 & 24
& 40.67 & 21.33 & 62
& 45.33 & 0.67 & 43
& 40.00 & 14.67 & 104
& 25.33 & 5.33 & 73
& 34.67 & 18.67 & 59
& 44.00 & 22.00 & 72
& 73.33 & 39.33 & 81
& 73.33 & 45.33 & 90 \\

PowerChain-AR Agent
& 9.33 & 0.00 & 137
& 12.67 & 2.00 & 36
& 78.00 & \textbf{57.33} & 76
& 66.00 & 3.33 & 44
& 88.00 & 31.33 & 104
& 55.33 & 9.33 & 46
& 68.00 & 12.67 & 76
& 74.67 & 21.33 & 84
& 91.33 & \textbf{67.33} & 78
& 92.67 & \textbf{68.00} & 66 \\

JIT-supervised Agent
& 6.00 & 2.00 & 170
& 5.33 & 4.67 & 28
& 75.33 & 28.67 & 94
& 68.67 & 18.00 & 70
& 76.00 & \textbf{32.00} & 139
& 56.00 & \textbf{20.00} & 63
& 83.33 & 41.33 & 109
& 80.67 & 27.33 & 127
& 93.33 & 48.00 & 89
& 96.00 & 43.33 & 64 \\

\PC{} Agent
& \textbf{12.67} & \textbf{4.00} & 250
& \textbf{16.67} & \textbf{8.67} & 48
& \textbf{88.00} & 45.33 & 90
& \textbf{89.33} & \textbf{30.00} & 76
& \textbf{92.67} & \textbf{32.00} & 135
& \textbf{73.33} & 14.67 & 64
& \textbf{90.67} & \textbf{43.33} & 94
& \textbf{90.67} & \textbf{29.33} & 121
& \textbf{97.33} & 54.67 & 90
& \textbf{98.00} & 47.33 & 74 \\
\bottomrule
\end{tabular}%
}

\vspace{0.6ex}
{\scriptsize
\raggedright
AR = Adaptive Retrieval. JIT = Just-in-Time. P@1 = Pass@1 and Pr = Precision over all 150 held-out queries. Tk = mean total tokens in thousands over successful runs out of the 150.
\par}
\end{minipage}
\end{table*}

\subsection{Pass@1 across Model Families}

\begin{figure*}[h]
    \centering
    \includegraphics[width=0.68\textwidth]{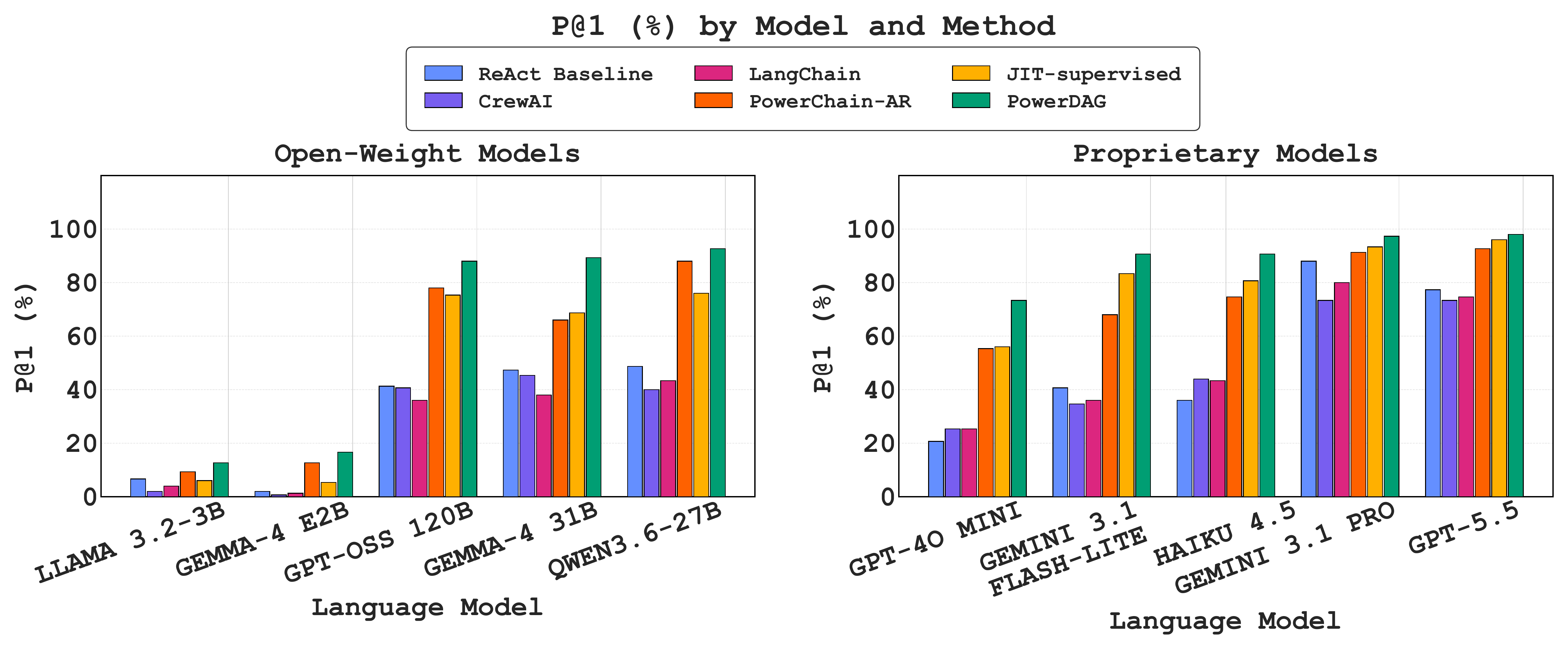}
    \caption{\textbf{Pass@1 by model and method.} The benchmark contains 150 queries across ten task families. \PC{} gives the highest Pass@1 for every evaluated LLM.}
    \label{fig:pass_precision}
\end{figure*}

\noindent \PC{} obtains the highest Pass@1 for every evaluated LLM. 
For proprietary models, \PC{} reaches $98.00\%$ on \llm{GPT-5.5}, $97.33\%$ on \llm{Gemini 3.1 Pro}, $90.67\%$ on \llm{Gemini 3.1 Flash-Lite} and \llm{Claude Haiku 4.5}, and $73.33\%$ on \llm{GPT-4o Mini}. 
For open-weight models, \PC{} reaches $92.67\%$ on \llm{Qwen3.6-27B}, $89.33\%$ on \llm{Gemma-4 31B}, $88.00\%$ on \llm{GPT-OSS 120B}, $16.67\%$ on \llm{Gemma-4 E2B}, and $12.67\%$ on \llm{Llama 3.2-3B}.
In comparison with the best non-\PC{} configuration for each model, \PC{} improves Pass@1 by $2.00\%$ to $20.66\%$.
%; we evaluate each model--method pair on 150 queries, so even the minimum $2.00$-point gain corresponds to three additional successful queries.
The largest gains occur on \llm{Gemma-4 31B} ($+20.66$), \llm{GPT-4o Mini} ($+17.33$), \llm{Claude Haiku 4.5} ($+10.00$), and \llm{GPT-OSS 120B} ($+10.00$).
Fig.~\ref{fig:pass_precision} shows Pass@1 for all six configurations across both model families.

\subsection{Precision, Token Use, and Advisories}
\noindent \textit{Precision:} \PC{} achieves the best or tied-best Precision on six of ten models.
PowerChain-AR achieves higher Precision on \llm{GPT-OSS 120B}, \llm{Gemini 3.1 Pro}, and \llm{GPT-5.5}.
The JIT-supervised agent achieves higher Precision on \llm{GPT-4o Mini}.
%\PC{} improves Pass@1 more consistently across models than Precision.
\noindent \textit{Token use and advisories:} We count token use over successful runs only, unless stated otherwise.
Across all 9,000 runs, including failed attempts, \PC{} uses 113.4k tokens per run on average, more than PowerChain-AR (86.5k), ReAct (84.8k), LangChain (85.0k), and CrewAI (79.0k), and comparable to the JIT-supervised agent (113.7k).
Retrieved exemplars and supervisor advisories add context to every prompt, which accounts for the higher per-run cost.
For high-capacity models, our results show that the agent uses fewer tokens than the no-exemplar baselines. For example, when using \llm{GPT-5.5}, \PC{} averages 73.95k tokens per successful run, compared with 84.04k for ReAct, 92.50k for LangChain, and 90.14k for CrewAI, because exemplars guide the agent to the correct workflow in fewer steps.
Restricting the comparison to queries that both systems solve yields the same ranking, showing that this lower per-query cost is not an artifact of differing success rates.
In 78.8\% of \PC{} runs, the JIT supervisor issues at least one advisory.
Only 0.7\% of runs with prerequisite advisories fail afterward.
All eight cases occur on open-weight models, mostly \llm{Gemma-4 E2B} and \llm{Llama 3.2-3B}, where the agent does not follow the repair implied by a correct advisory.
Appendix~\ref{appx:supervisor_diagnostics} reports rule coverage, advisory rate, blocked-call rate, and the most frequently blocked tools.

% {\color{blue}
\subsection{Effect of Combining Retrieval and Supervision}
\noindent PowerChain-AR uses adaptive retrieval without JIT supervision. 
The JIT-supervised agent uses prerequisite checks without workflow exemplars. 
\PC{} combines both mechanisms and gives a higher Pass@1 than both ablations on every model.
Appendix~\ref{appx:stage_selection} (Table~\ref{tab:stage1_cutoff_comparison}) shows that the adaptive cutoff outperforms any fixed top-$k$ in retrieval quality.
The gain over PowerChain-AR ranges from $3.34\%$ on \llm{Llama 3.2-3B} to $23.33\%$ on \llm{Gemma-4 31B}. 
The gain over the JIT-supervised agent ranges from $2.00\%$ on \llm{GPT-5.5} to $20.66\%$ on \llm{Gemma-4 31B}. 
Exemplar retrieval and prerequisite checking address distinct failure modes.
Exemplars improve tool selection and call ordering, while the JIT supervisor prevents execution when prerequisites are unmet.
Combining both is necessary to achieve the highest Pass@1.

Overall, \PC{} achieves the highest Pass@1 on every evaluated LLM and the highest Precision on six of ten models. 
For the four remaining models, the configuration that achieves higher Precision has a lower Pass@1 than \PC{}.
% }

% \section{Discussion}
% \input{discussion}

\section{Conclusion}
\label{sec:conclusion}

\noindent Two gaps limit prior agentic approaches to distribution-grid analysis: (i) fixed-size context retrieval that ignores query complexity, and (ii) silent tool failures from unmet prerequisites.
\PC{} addresses both with adaptive exemplar retrieval and JIT supervision.
We draw five conclusions from the benchmark results.
\begin{itemize}
    \item {\textit{Correctness.} \PC{} achieves the highest Pass@1 on every evaluated LLM. 
    \llm{GPT-5.5} reaches $98.00\%$, \llm{Gemini 3.1 Pro} reaches $97.33\%$, and \llm{Qwen3.6-27B} reaches $92.67\%$.} \PC{} improves over the strongest non-\PC{} configuration by $2.00\%$ to $20.66\%$.

    \item {\textit{Complementary components.} Our ablation study confirms that each component addresses a distinct failure mode. Removing retrieval (the JIT-supervised agent) reduces Pass@1 by up to $20.66\%$, and removing supervision (PowerChain-AR) reduces it by up to $23.33\%$. Both components are necessary to achieve the highest Pass@1 on every model.}

    \item {\textit{Token cost.} \PC{} uses more tokens per run on average (113.4k) than the no-exemplar baselines (82.7k) because exemplars and advisories add context to every prompt. 
    For the strongest models, successful \PC{} runs also use fewer tokens on average than successful baseline runs.
    This suggests that, when the model can follow the retrieved workflow, \PC{} reduces error-driven extra calls.}
    %because exemplars reduce the steps needed to reach a correct workflow.}

    \item {\textit{Open-weight models.} Strong open-weight models demonstrate competitive performance. 
    \llm{Qwen3.6-27B} reaches $92.67\%$ Pass@1, \llm{Gemma-4 31B} reaches $89.33\%$, and \llm{GPT-OSS 120B} reaches $88.00\%$. This supports local deployment for utilities that cannot share grid data with commercial APIs. Small models fall below a capability threshold. \llm{Gemma-4 E2B} reaches $16.67\%$ and \llm{Llama 3.2-3B} reaches $12.67\%$.}

    \item {\textit{Benchmark.} We release 200 expert-verified query-workflow records across ten distribution-grid task families, an exemplar archive, a stateful execution environment, and an evaluator as a reproducible benchmark for agentic workflow correctness.}

\end{itemize}

% \noindent A natural next step is to close the loop with grid operations. \PC{} now produces validated studies offline, but deploying it on the live grid requires the supervisor to certify that each recommended action, such as a curtailment, dispatch, or switching command, is feasible and within operating limits before it reaches equipment, and to re-plan as feeder conditions change. Certifying agent actions on physical infrastructure, not only their tool order, is the central barrier to operational deployment.

%{\color{blue}
%\noindent We evaluate \PC{} on well-formed expert queries under a within-family held-out split.
%Future work will test robustness to malformed and out-of-family queries, measure engineer time savings, quantify supervisor false-positive and false-negative rates, and validate that numerical outputs are physically correct, not only that the required tools were called in the right order.
%}

%\section{AI Usage Disclosure}
\label{sec:disclosure}
% \noindent \textbf{AI Usage Disclosure:} We used generative AI to revise grammar and improve writing.
% We reviewed each suggestion and take responsibility for the final content.

% Appendix - uses same font size as references (normalsize)
\setlist[enumerate]{itemsep=0pt, parsep=0pt, topsep=3pt, partopsep=0pt, leftmargin=*, label=\arabic*.}

\begin{appendices}
\footnotesize

\counterwithin{equation}{section}

% \noindent \textbf{AI Usage Disclosure:} We used generative AI to revise grammar and improve writing.
% We reviewed each suggestion and took responsibility for the final content.

% Appendix
\section{Benchmark Task Set}
\label{appx:queries}

\noindent We curate 200 query-workflow records across ten distribution-grid task families, with 20 records per family (Table~\ref{tab:benchmark_families}).
Each record consists of a natural-language query and its corresponding tool-call workflow sequence.
We validate each record by executing the workflow in the evaluation environment and having a domain expert verify that the tool-call sequence, argument bindings, and outputs are correct.
We partition the 200 records into 50 exemplars (5 per family), which serve as the in-context archive available to the agent, and 150 held-out evaluation queries (15 per family), which we withhold from all stages of agent development and use as the test set for all reported metrics.
The full dataset is available at \url{https://github.com/emmanuelbadmus/DistGrid-AgentBench}.

\begin{table}[htbp]
\centering
\caption{Benchmark task families and data split.}
\label{tab:benchmark_families}
\scriptsize
\setlength{\tabcolsep}{3pt}
\begin{tabular}{p{0.20\linewidth} p{0.52\linewidth} c c}
\toprule
\textbf{Family} & \textbf{Workflow scope} & \textbf{Total} & \textbf{Held-out} \\
\midrule
General & inventory, metadata lookup, node export & 20 & 15 \\
Powerflow & feeder power-flow solve and validation & 20 & 15 \\
Infeasibility & $\ell_1/\ell_2$ current-slack diagnostics & 20 & 15 \\
DHC & hosting-capacity and curtailment studies & 20 & 15 \\
EV & charger-candidate screening and placement & 20 & 15 \\
BESS & sizing, dispatch, tariff, degradation, and economics & 20 & 15 \\
PV & PV irradiance, parameters, and generation & 20 & 15 \\
GFI & inverter control, disturbance, and stability analysis & 20 & 15 \\
Combined T\&D & coupled transmission-distribution workflows & 20 & 15 \\
DSSE & state estimation and bad-measurement detection & 20 & 15 \\
\bottomrule
\end{tabular}

\vspace{0.5ex}
{\scriptsize Each family has 20 records: 5 exemplar records and 15 held-out evaluation records.}
\end{table}

% \refstepcounter{section}
\section{Annotated Workflow Exemplars}
\label{appx:awf}

\noindent We reserve 50 of the 200 records as the exemplar archive, five per task family.
Each exemplar is a query-workflow record $(q^{(i)}, w^{(i)})$ that pairs a natural-language query $q^{(i)}$ with its validated tool-call workflow $w^{(i)}$.
The agent retrieves a relevant set from this archive at inference time.
No held-out evaluation query appears in the archive.
% The full exemplar archive is available at \url{https://github.com/emmanuelbadmus/DistGrid-AgentBench}.

\begin{PromptBox}[title={\textbf{Example Query-Workflow Record}}]
\{ "id": "001", "query": "How many capacitors are in the Rochester feeder?",\\
\quad "workflow": [\{"name":"load\_distribution\_network", \ldots\}, \{"name":"get\_component\_count", \ldots\}] \}\\[0.15ex]
\(\vdots\)\\[0.15ex]
\{ "id": "007", "query": "Run steady-state three-phase power flow on Stowe for 2025-03-21 09:00, then plot bus voltage magnitudes.",\\
\quad "workflow": [\\
\quad\quad \{"name": "load\_load", "arguments": \{"feeder": "stowe", "timestamp": "2025-03-21 09:00:00"\}\},\\
\quad\quad \{"name": "load\_solar", "arguments": \{"feeder": "stowe", "timestamp": "2025-03-21 09:00:00"\}\},\\
\quad\quad \{"name": "load\_distribution\_network", "arguments": \{"feeder": "stowe"\}\},\\
\quad\quad \{"name": "create\_and\_initialize\_model", "arguments": \{\}\},\\
\quad\quad \{"name": "build\_constraints", "arguments": \{"analysis\_type": "powerflow"\}\},\\
\quad\quad \{"name": "build\_objective", "arguments": \{"analysis\_type": "powerflow"\}\},\\
\quad\quad \{"name": "solve", "arguments": \{\}\},\\
\quad\quad \{"name": "update\_network\_voltages", "arguments": \{\}\},\\
\quad\quad \{"name": "plot\_network\_data", "arguments": \{"plot\_type": "voltage", "feeder": "stowe"\}\}\\
\quad ] \}\\[0.15ex]
\(\vdots\)\\[0.15ex]
\{ "id": "050", \ldots \}
\end{PromptBox}

\section{Retrieval Diagnostics}
\label{appx:stage_selection}

\noindent Let $\mathcal{Q}_{\mathrm{eval}}$ denote the 150 held-out queries and $\mathcal{Q}_{\mathrm{archive}}=\{q_i\}_{i=1}^{50}$ the 50 archived exemplar queries.
For each query $q\in\mathcal{Q}_{\mathrm{eval}}$, Stage~1 ranks all queries in $\mathcal{Q}_{\mathrm{archive}}$.
A retrieved exemplar is \emph{relevant} if it belongs to the same task family as $q$.
Because the archive contains five exemplars per family, each evaluation query has five relevant exemplars.
We report four standard IR metrics~\cite{manning2008introduction,jarvelin2002cumulated}.

\noindent\textbf{MRR}~\cite{manning2008introduction}: Let $r_q$ denote the rank of the first relevant exemplar retrieved for query $q$.
MRR averages its reciprocal rank over all evaluation queries, rewarding scorers that place at least one same-family exemplar near the top:
\[
\mathrm{MRR}
=
\frac{1}{|\mathcal{Q}_{\mathrm{eval}}|}
\sum_{q\in\mathcal{Q}_{\mathrm{eval}}}
\frac{1}{r_q}
\]

\noindent\textbf{MAP}~\cite{manning2008introduction}: For each query $q$, Average Precision (AP) is the mean precision at ranks containing relevant exemplars.
MAP averages AP over all evaluation queries:
\[
\mathrm{MAP}
=
\frac{1}{|\mathcal{Q}_{\mathrm{eval}}|}
\sum_{q\in\mathcal{Q}_{\mathrm{eval}}}
\frac{1}{|\mathcal{R}_q|}
\sum_{k:\,\mathrm{rel}_{q,k}=1}
\mathrm{Pr}_q@k
\]
where $\mathcal{R}_q$ is the set of relevant archived exemplars for $q$, $|\mathcal{R}_q|=5$, $\mathrm{rel}_{q,k}$ indicates whether the exemplar at rank $k$ is relevant, and $\mathrm{Pr}_q@k$ is precision through rank $k$.

\noindent\textbf{nDCG@5}~\cite{jarvelin2002cumulated}: Normalized Discounted Cumulative Gain rewards relevant exemplars appearing near the top:
\[
\mathrm{nDCG@5}
=
\frac{1}{|\mathcal{Q}_{\mathrm{eval}}|}
\sum_{q\in\mathcal{Q}_{\mathrm{eval}}}
\frac{\mathrm{DCG}_q@5}{\mathrm{IDCG}_q@5},
\; 
\mathrm{DCG}_q@5
=
\sum_{k=1}^{5}
\frac{\mathrm{rel}_{q,k}}{\log_2(k+1)}
\]
Here, $k$ is retrieval rank, $\mathrm{rel}_{q,k}=1$ if the exemplar at rank $k$ is relevant to $q$ and $0$ otherwise, and $\mathrm{IDCG}_q@5$ is the maximum possible $\mathrm{DCG}_q@5$, obtained when all five relevant exemplars occupy top-five positions.

\noindent\textbf{Precision@5}~\cite{manning2008introduction}: Precision@5 is the average fraction of the top five retrieved exemplars that are relevant:
\[
\mathrm{Precision@5}
=
\frac{1}{|\mathcal{Q}_{\mathrm{eval}}|}
\sum_{q\in\mathcal{Q}_{\mathrm{eval}}}
\frac{1}{5}\sum_{k=1}^{5}\mathrm{rel}_{q,k}
\]
Table~\ref{tab:stage1_scorer_comparison} reports these metrics.
Google \llm{gemini-embedding-2} with cosine similarity gives the best performance, leading three of four metrics.

\begin{table}[H]
\centering
\caption{Stage 1 scorer comparison for retrieval quality.}
\label{tab:stage1_scorer_comparison}
\tiny
\setlength{\tabcolsep}{2pt}
\resizebox{\linewidth}{!}{
\begin{tabular}{lcccc}
\toprule
\textbf{Scorer} & \textbf{MRR} & \textbf{MAP} & \textbf{nDCG@5} & \textbf{Precision@5} \\
\midrule
BM25 & 0.914 & 0.727 & 0.709 & 0.660 \\
TF-IDF + cosine & 0.937 & 0.772 & 0.736 & 0.680 \\
char n-gram TF-IDF + cosine & 0.943 & 0.783 & 0.757 & 0.704 \\
LSA/SVD + cosine & \textbf{0.948} & 0.780 & 0.767 & 0.715 \\
OpenAI text-embedding-3-small + cosine & 0.932 & 0.766 & 0.740 & 0.681 \\
OpenAI text-embedding-3-large + cosine & 0.918 & 0.765 & 0.743 & 0.695 \\
Google gemini-embedding-001 + cosine & 0.935 & 0.802 & 0.775 & 0.731 \\
Google gemini-embedding-2 + cosine & 0.937 & \textbf{0.809} & \textbf{0.791} & \textbf{0.755} \\
% ColBERT-style MaxSim & 0.909 & 0.676 & 0.658 & 0.593 \\
% cross-encoder pair score & 0.743 & 0.539 & 0.504 & 0.471 \\
all-MiniLM-L6-v2 + cosine & 0.924 & 0.738 & 0.712 & 0.653 \\
all-mpnet-base-v2 + cosine & 0.933 & 0.756 & 0.729 & 0.673 \\
\bottomrule
\end{tabular}
}
\end{table}

\noindent Using Google \llm{gemini-embedding-2}, we compare fixed top-$k$ retrieval against five adaptive cutoff policies in Table~\ref{tab:stage1_cutoff_comparison}.
Avg.~Count is the average number of Stage 1 candidates per query.
F1 is the harmonic mean of precision (the fraction of candidates from the same family as the query) and recall (the fraction of the five same-family exemplars included in the candidate set).
Let $\mathrm{FR}$ denote the fraction of queries for which all five same-family exemplars are retrieved.
RCS is the harmonic mean of F1 and $\mathrm{FR}$:
\begin{equation}
\mathrm{RCS}
=
\frac{2\cdot\mathrm{F1}\cdot\mathrm{FR}}
{\mathrm{F1}+\mathrm{FR}}.
\end{equation}
The adaptive two-segment elbow cutoff achieves the highest RCS (0.635), giving the best balance between candidate precision and exemplar recovery.

\begin{table}[H]
\centering
\caption{Stage 1 cutoff-policy comparison using the selected Gemini-2 scorer.}
\label{tab:stage1_cutoff_comparison}
\scriptsize
\setlength{\tabcolsep}{2pt}
\resizebox{\linewidth}{!}{
\begin{tabular}{lrrrr}
\toprule
\textbf{Policy}
& \multicolumn{1}{c}{\textbf{Avg. Count}}
& \multicolumn{1}{c}{\textbf{F1}}
& \multicolumn{1}{c}{\textbf{Full Rec.}}
& \multicolumn{1}{c}{\textbf{RCS}} \\
\midrule
top\_5  & 5.000  & \textbf{0.755} & 0.353 & 0.481 \\
top\_50 & 50.000 & 0.182 & \textbf{1.000} & 0.308 \\
\midrule
adaptive\_largest\_gap         & 14.060 & 0.502 & 0.447 & 0.473 \\
adaptive\_two\_segment\_elbow  & 14.407 & 0.623 & 0.647 & \textbf{0.635} \\
adaptive\_two\_segment\_bic    & 13.673 & 0.615 & 0.627 & 0.621 \\
adaptive\_otsu\_separation     & 21.580 & 0.476 & 0.887 & 0.619 \\
adaptive\_kneedle\_distance    & 13.100 & 0.595 & 0.627 & 0.611 \\
\bottomrule
\end{tabular}
}
\end{table}

% \noindent Table~\ref{tab:stage2_filter_comparison} runs the same fixed top-$k$ baselines and the adaptive elbow through the Stage 2 filter.

% \rev{\begin{table}[H]
% \centering
% \caption{Stage 2 workflow-injection results using the selected Gemini-2 Stage 1 scorer.}
% \label{tab:stage2_filter_comparison}
% \scriptsize
% \setlength{\tabcolsep}{2pt}
% \resizebox{\linewidth}{!}{
% \begin{tabular}{lrrrrrr}
% \toprule
% \textbf{Policy} & \textbf{Injected} & \textbf{Prec.} & \textbf{Recall} & \textbf{F1} & \textbf{Domin.} & \textbf{Cand. Tokens} \\
% \midrule
% top\_5 & \textbf{1.900} & 0.858 & 0.340 & 0.470 & \textbf{0.850} & \textbf{622.500} \\
% % top\_25 & 3.000 & 0.760 & 0.460 & 0.565 & 0.750 & 3365.100 \\
% top\_50 & 2.850 & 0.792 & \textbf{0.470} & \textbf{0.579} & 0.750 & 6266.300 \\
% adaptive\_two\_segment\_elbow & 2.500 & \textbf{0.860} & 0.420 & 0.540 & \textbf{0.850} & 2280.800 \\
% \bottomrule
% \end{tabular}
% }
% \end{table}}
% \noindent The adaptive elbow matches top-5 on injected-set precision ($0.860$) and dominance ($0.850$) while recovering more exemplars (recall $0.420$ versus $0.340$), at roughly one-third of top-50's candidate tokens.

\section{Derivation of Adaptive Two-Segment Elbow Cutoff}
\label{appx:cutoff}

\noindent We derive the \texttt{adaptive\_two\_segment\_elbow} cutoff objective used in \eqref{eq:elbow}.
We define $\Phi$ as the segment fitting error.
Assume $\{\mathrm{Sim}_i\}_{i=1}^N$ is sorted in non-increasing order as in \eqref{eq:sim_rank}.
For a candidate breakpoint $b\in\{2,\ldots,N-2\}$, define the left and right index sets:
\begin{equation}
I_L(b) := \{1,\ldots,b\}, \qquad I_R(b) := \{b+1,\ldots,N\}
\label{eq:app_segments}
\end{equation}

\noindent \textbf{Least-squares line on an index interval.}
For any interval $r{:}s$ with $1\le r<s\le N$, define the best affine fit of $(i,\mathrm{Sim}_i)$ by
\begin{equation}
(\hat\lambda_{r:s},\hat\mu_{r:s})
:=\arg\min_{\lambda,\mu\in\mathbb{R}}
\sum_{i=r}^{s}\Big(\mathrm{Sim}_i-(\lambda+\mu i)\Big)^2
\label{eq:app_ls}
\end{equation}

\noindent \textbf{Segment Error $\Phi(r,s)$.}
We define $\Phi(r,s)$ as the RMSE of the least-squares affine fit to $\{(i,\mathrm{Sim}_i)\}_{i=r}^{s}$:
\begin{equation}
\Phi(r,s)
=\sqrt{\frac{1}{s-r+1}\sum_{i=r}^{s}\Big(\mathrm{Sim}_i-(\hat\lambda_{r:s}+\hat\mu_{r:s}i)\Big)^2 }
\label{eq:app_phi}
\end{equation}

\noindent \textbf{Weighted two-segment objective.}
For breakpoint $b$, we combine the two segment RMSE values as
\begin{align}
\mathcal{L}_{\mathrm{elbow}}(b)
&:=\frac{|I_L(b)|}{N}\,\Phi(1,b)+\frac{|I_R(b)|}{N}\,\Phi(b+1,N) \nonumber\\
&=\frac{b}{N}\Phi(1,b)+\frac{N-b}{N}\Phi(b+1,N)
\label{eq:app_energy}
\end{align}

\section{Workflow-Filter Prompt}
\label{app:filter_prompt}

\noindent We filter the Stage 1 candidate set $W_{\mathrm{cand}}$ with the following LLM prompt.
At inference time, the system replaces \texttt{\{query\} } with the unseen query $q$ and \texttt{\{candidates\} } with the JSON-formatted candidate records.

\begin{PromptBox}[title={\textbf{Workflow Filter Prompt}}]
Select the best out of these candidate workflows to keep as in-context exemplars.

USER QUERY: \{query\}

CANDIDATES (JSON): \{candidates\}

Each candidate has:
- "query"
- "workflow" (tool name + arguments)

Keep a candidate if its workflow helps solve the user query.
Exclude only if clearly unrelated.

Return ONLY a JSON list of indices to keep (no duplicates).
Example: [0, 2, 5]
\end{PromptBox}

% ==========================================================

\section{Supervisor Advisory and Diagnostics}
\label{appx:supervisor_diagnostics}

\noindent Before each tool call, the supervisor checks whether the environment satisfies the prerequisite states of the call.
If a prerequisite is missing, the supervisor blocks the call, returns an advisory to the agent, and leaves the environment state unchanged.

\begin{PromptBox}[title={\textbf{Advisory Template}}]
Error: Supervisor blocked \texttt{\{tool\_name\} } before execution. \texttt{\{violation\_msg\} }\\[0.5ex]
Call the missing prerequisite tool first.
Retry the blocked tool only if it is still needed.\\[0.5ex]
Do not restart the workflow.
Do not repeat successful loader or setup tools.
\end{PromptBox}

% \noindent We build the prerequisite rule library from expert workflow traces and tool read/write annotations.
% Table~\ref{tab:supervisor_coverage} reports its coverage: the supervisor checks 94 of 108 registered tools and 1,097 of 1,330 held-out expert tool calls.

\noindent We build the supervisor's prerequisite rules from expert-verified workflows and tool-level state contracts that specify which tools create, modify, or require environment state.
Table~\ref{tab:supervisor_coverage} reports two forms of coverage. For tool coverage, 94 of 108 registered tools have explicit prerequisite checks. For call coverage, these checks cover 1,097 of 1,330 tool calls in the held-out expert workflows.

\begin{table}[H]
\centering
\caption{Supervisor rule coverage.}
\label{tab:supervisor_coverage}
\scriptsize
\setlength{\tabcolsep}{3pt}
\begin{tabular*}{\columnwidth}{@{\extracolsep{\fill}} l r l r @{}}
\toprule
\textbf{Metric} & \textbf{Value} & \textbf{Metric} & \textbf{Value} \\
\midrule
Registered tools & 108 & Tools in $\mathrm{dom}(C)$ & 94 (87.0\%) \\
Held-out expert calls & 1,330 & Calls to $\mathrm{dom}(C)$ tools & 1,097 (82.5\%) \\
Distinct held-out tools & 103 & Distinct tools in $\mathrm{dom}(C)$ & 90 (87.4\%) \\
Prerequisite rules & 112 & Post-advisory failures & 8 / 1,178 (0.7\%) \\
\bottomrule
\end{tabular*}
\end{table}

\noindent The rules span all ten task families in Table~\ref{tab:benchmark_families}.
In full \PC{} runs, the supervisor triggers 109 distinct rules, issues advisories in 78.8\% of runs, and blocks 6.57 proposed calls per run on average.
In the JIT-only ablation, the supervisor issues advisories in 87.5\% of runs and blocks 4.64 calls per run.
A rule-level audit finds no supervisor misclassifications, with $0.0\%$ false positives on 1,034 admissible calls and $0.0\%$ false negatives on 1,393 constructed prerequisite violations. The remaining failures are agent-compliance errors, where 8 of 1,178 advisory runs fail afterward, all on smaller models.

% Algorithm~\ref{lst:dag_equiv_short} shows the routine that tests call independence and normalizes a trace by swapping adjacent independent calls into a fixed order before scoring.

% \begin{lstlisting}[language=Python, caption={Order-invariant scoring via independence checks and trace normalization.}, label={lst:dag_equiv_short}]
% def are_independent(t1, t2):
%     r1, w1 = get_rw_sets(t1)   # abstract read/write sets
%     r2, w2 = get_rw_sets(t2)

%     if "ALL" in w1 or "ALL" in w2:
%         return False
%     if not w1.isdisjoint(w2):  # write-write conflict
%         return False
%     if not w1.isdisjoint(r2):  # write-read conflict
%         return False
%     if not w2.isdisjoint(r1):  # read-write conflict
%         return False
%     return True

% def normalize_trace(trace):
%     trace = list(trace)
%     for _ in range(len(trace)):
%         swapped = False
%         for j in range(len(trace) - 1):
%             if (are_independent(trace[j], trace[j+1])
%                 and sort_key(trace[j]) > sort_key(trace[j+1])):
%                 trace[j], trace[j+1] = trace[j+1], trace[j]
%                 swapped = True
%         if not swapped:
%             break
%     return trace
% \end{lstlisting}

% ==========================================================
\end{appendices}

\bibliographystyle{IEEEtran}
\bibliography{references}

@IEEEtranBSTCTL{BSTcontrol,
  CTLuse_forced_etal       = "yes",
  CTLmax_names_forced_etal = "2",
  CTLnames_show_etal       = "1"
}

@article{chen2021evaluating,
  title={Evaluating large language models trained on code},
  author={Chen, Mark and Tworek, Jerry and Jun, Heewoo and Yuan, Qiming and Pinto, Henrique Ponde De Oliveira and Kaplan, Jared and Edwards, Harri and Burda, Yuri and Joseph, Nicholas and Brockman, Greg and others},
  journal={arXiv preprint arXiv:2107.03374},
  year={2021}
}

@misc{anthropic2026advisor,
  author       = {{Anthropic}},
  title        = {Advisor Tool},
  year         = {2026},
  howpublished = {\url{https://platform.claude.com/docs/en/agents-and-tools/tool-use/advisor-tool}},
  note         = {Accessed: 2026-06-22}
}

@misc{google_gemini_embedding2_2026,
  title        = {Gemini Embedding 2},
  howpublished = {Google AI for Developers Documentation},
  note         = {Accessed 2026-05-16},
  url          = {https://ai.google.dev/gemini-api/docs/embeddings}
}

@inproceedings{bank2013analysis,
  title={Analysis of the impacts of distribution connected PV using high-speed datasets},
  author={Bank, Jason and Mather, Barry},
  booktitle={2013 IEEE Green Technologies Conference (GreenTech)},
  pages={153--159},
  year={2013},
  organization={IEEE}
}

@inproceedings{badmus2024anoca,
  title={Anoca: Ac network-aware optimal curtailment approach for dynamic hosting capacity},
  author={Badmus, Emmanuel O and Pandey, Amritanshu},
  booktitle={2024 IEEE 63rd Conference on Decision and Control (CDC)},
  pages={5338--5345},
  year={2024},
  organization={IEEE}
}

@inproceedings{bhattaram2025geoflow,
  title={GeoFlow: Agentic Workflow Automation for Geospatial Tasks},
  author={Bhattaram, Amulya and Chung, Justin and Chung, Stanley and Gupta, Ranit and Ramamoorthy, Janani and Gullapalli, Kartikeya and Marculescu, Diana and Stamoulis, Dimitrios},
  booktitle={Proceedings of the 33rd ACM International Conference on Advances in Geographic Information Systems},
  pages={1150--1153},
  year={2025}
}

@article{qin2023toolllm,
  title={Toolllm: Facilitating large language models to master 16000+ real-world apis},
  author={Qin, Yujia and Liang, Shihao and Ye, Yining and Zhu, Kunlun and Yan, Lan and Lu, Yaxi and Lin, Yankai and Cong, Xin and Tang, Xiangru and Qian, Bill and others},
  journal={arXiv preprint arXiv:2307.16789},
  year={2023}
}

@inproceedings{yao2022react,
  title={React: Synergizing reasoning and acting in language models},
  author={Yao, Shunyu and Zhao, Jeffrey and Yu, Dian and Du, Nan and Shafran, Izhak and Narasimhan, Karthik R and Cao, Yuan},
  booktitle={The eleventh international conference on learning representations},
  year={2022}
}

@article{lewis2020retrieval,
  title={Retrieval-augmented generation for knowledge-intensive nlp tasks},
  author={Lewis, Patrick and Perez, Ethan and Piktus, Aleksandra and Petroni, Fabio and Karpukhin, Vladimir and Goyal, Naman and K{\"u}ttler, Heinrich and Lewis, Mike and Yih, Wen-tau and Rockt{\"a}schel, Tim and others},
  journal={Advances in neural information processing systems},
  volume={33},
  pages={9459--9474},
  year={2020}
}

@article{brown2020language,
  title={Language models are few-shot learners},
  author={Brown, Tom and Mann, Benjamin and Ryder, Nick and Subbiah, Melanie and Kaplan, Jared D and Dhariwal, Prafulla and Neelakantan, Arvind and Shyam, Pranav and Sastry, Girish and Askell, Amanda and others},
  journal={Advances in neural information processing systems},
  volume={33},
  pages={1877--1901},
  year={2020}
}

@article{wu2024avatar,
  title={Avatar: Optimizing llm agents for tool usage via contrastive reasoning},
  author={Wu, Shirley and Zhao, Shiyu and Huang, Qian and Huang, Kexin and Yasunaga, Michihiro and Cao, Kaidi and Ioannidis, Vassilis N and Subbian, Karthik and Leskovec, Jure and Zou, James},
  journal={Advances in Neural Information Processing Systems},
  volume={37},
  pages={25981--26010},
  year={2024}
}

@incollection{kersting2018distribution,
  title={Distribution system modeling and analysis},
  author={Kersting, William H},
  booktitle={Electric power generation, transmission, and distribution},
  pages={26--1},
  year={2018},
  publisher={CRC press}
}

@book{buchanan1984rule,
  title={Rule based expert systems: the mycin experiments of the stanford heuristic programming project (the Addison-Wesley series in artificial intelligence)},
  author={Buchanan, Bruce G and Shortliffe, Edward H},
  year={1984},
  publisher={Addison-Wesley Longman Publishing Co., Inc.}
}

@techreport{ewab2025opportunities,
  title={Opportunities for American Workers in Energy},
  author={{21st Century Energy Workforce Advisory Board}},
  institution={U.S. Department of Energy},
  year={2025},
  month={jul},
  url={https://www.energy.gov/sites/default/files/2025-07/EWAB_Special_Report_Opportunities_for_American_Workers_in_Energy.pdf}
}

@article{patil2024gorilla,
  title={Gorilla: Large language model connected with massive apis},
  author={Patil, Shishir G and Zhang, Tianjun and Wang, Xin and Gonzalez, Joseph E},
  journal={Advances in Neural Information Processing Systems},
  volume={37},
  pages={126544--126565},
  year={2024}
}

@article{liu2024lost,
  title={Lost in the middle: How language models use long contexts},
  author={Liu, Nelson F and Lin, Kevin and Hewitt, John and Paranjape, Ashwin and Bevilacqua, Michele and Petroni, Fabio and Liang, Percy},
  journal={Transactions of the association for computational linguistics},
  volume={12},
  pages={157--173},
  year={2024}
}

@inproceedings{min2022rethinking,
  title={Rethinking the role of demonstrations: What makes in-context learning work?},
  author={Min, Sewon and Lyu, Xinxi and Holtzman, Ari and Artetxe, Mikel and Lewis, Mike and Hajishirzi, Hannaneh and Zettlemoyer, Luke},
  booktitle={Proceedings of the 2022 conference on empirical methods in natural language processing},
  pages={11048--11064},
  year={2022}
}

@article{schick2023toolformer,
  title={Toolformer: Language models can teach themselves to use tools},
  author={Schick, Timo and Dwivedi-Yu, Jane and Dess{\`\i}, Roberto and Raileanu, Roberta and Lomeli, Maria and Hambro, Eric and Zettlemoyer, Luke and Cancedda, Nicola and Scialom, Thomas},
  journal={Advances in neural information processing systems},
  volume={36},
  pages={68539--68551},
  year={2023}
}

@article{shinn2023reflexion,
  title={Reflexion: Language agents with verbal reinforcement learning},
  author={Shinn, Noah and Cassano, Federico and Gopinath, Ashwin and Narasimhan, Karthik and Yao, Shunyu},
  journal={Advances in neural information processing systems},
  volume={36},
  pages={8634--8652},
  year={2023}
}

@article{wang2025agentspec,
  title={Agentspec: Customizable runtime enforcement for safe and reliable llm agents},
  author={Wang, Haoyu and Poskitt, Christopher M and Sun, Jun},
  journal={arXiv preprint arXiv:2503.18666},
  year={2025}
}

@inproceedings{jin2025gridmind,
  title={GridMind: LLMs-powered agents for power system analysis and operations},
  author={Jin, Hongwei and Kim, Kibaek and Kwon, Jonghwan},
  booktitle={Proceedings of the SC'25 Workshops of the International Conference for High Performance Computing, Networking, Storage and Analysis},
  pages={560--568},
  year={2025}
}

@article{zhang2025grid,
  title={Grid-agent: An LLM-powered multi-agent system for power grid control},
  author={Zhang, Yan and Saber, Ahmad Mohammad and Youssef, Amr and Kundur, Deepa},
  journal={arXiv preprint arXiv:2508.05702},
  year={2025}
}

@article{huang2024survey,
  title={A survey on retrieval-augmented text generation for large language models},
  author={Huang, Yizheng and Huang, Jimmy Xiangji},
  journal={ACM Computing Surveys},
  year={2024},
  publisher={ACM New York, NY}
}

@article{chen2025x,
  title={X-GridAgent: An LLM-Powered Agentic AI System for Assisting Power Grid Analysis},
  author={Chen, Xin and others},
  journal={arXiv preprint arXiv:2512.20789},
  year={2025}
}

@article{Luo2023DrICLDI,
  title={{Dr.ICL}: Demonstration-Retrieved In-context Learning},
  author={Luo, Man and Xu, Xin and Dai, Zhuyun and Pasupat, Panupong and Kazemi, Mehran and Baral, Chitta and Imbrasaite, Vaiva and Zhao, Vincent},
  journal={arXiv preprint arXiv:2305.14128},
  year={2023},
  url={https://arxiv.org/abs/2305.14128}
}

@inproceedings{xu2024enhancing,
  title={Enhancing tool retrieval with iterative feedback from large language models},
  author={Xu, Qiancheng and Li, Yongqi and Xia, Heming and Li, Wenjie},
  booktitle={Findings of the Association for Computational Linguistics: EMNLP 2024},
  pages={9609--9619},
  year={2024}
}

@article{wang2024agent,
  title={Agent workflow memory},
  author={Wang, Zora Zhiruo and Mao, Jiayuan and Fried, Daniel and Neubig, Graham},
  journal={arXiv preprint arXiv:2409.07429},
  year={2024}
}

@inproceedings{tan2025meta,
  title={Meta-agent-workflow: Streamlining tool usage in llms through workflow construction, retrieval, and refinement},
  author={Tan, Xiaoyu and Li, Bin and Qiu, Xihe and Qu, Chao and Chu, Wei and Xu, Yinghui and Qi, Yuan},
  booktitle={Companion Proceedings of the ACM on Web Conference 2025},
  pages={458--467},
  year={2025}
}

@article{yao2024taubench,
  title={$\tau$-bench: A Benchmark for Tool-Agent-User Interaction in Real-World Domains},
  author={Yao, Shunyu and Shinn, Noah and Razavi, Pedram and Narasimhan, Karthik},
  journal={arXiv preprint arXiv:2406.12045},
  year={2024}
}

@article{bonadia2023potential,
  title={On the potential of ChatGPT to generate distribution systems for load flow studies using OpenDSS},
  author={Bonadia, Rodrigo S and Trindade, Fernanda CL and Freitas, Walmir and Venkatesh, Bala},
  journal={IEEE Transactions on Power Systems},
  volume={38},
  number={6},
  pages={5965--5968},
  year={2023},
  publisher={IEEE}
}

@article{chassin2014gridlab,
  title={GridLAB-D: an agent-based simulation framework for smart grids},
  author={Chassin, David P and Fuller, Jason C and Djilali, Ned},
  journal={Journal of Applied Mathematics},
  volume={2014},
  number={1},
  pages={492320},
  year={2014},
  publisher={Wiley Online Library}
}

@article{liu2025repower,
  title={RePower: An LLM-driven autonomous platform for power system data-guided research},
  author={Liu, Yu-Xiao and Jia, Mengshuo and Zhang, Yong-Xin and Wang, Jianxiao and He, Guannan and Zhong, Shao-Long and Dang, Zhi-Min},
  journal={Patterns},
  volume={6},
  number={4},
  year={2025},
  publisher={Elsevier}
}

@article{she2026pfagent,
  title={PFAgent: A Tractable and Self-Evolving Power-Flow Agent for Interactive Grid Analysis},
  author={She, Buxin and Chen, Brian and Guo, Luanzheng and Li, Fangxing},
  journal={arXiv preprint arXiv:2604.10846},
  year={2026}
}

@article{liu2026grid,
  title={Grid-Orch: An LLM-Powered Orchestrator for Distribution Grid Simulation and Analytics},
  author={Liu, Boming and Dong, Jin and Lian, Jamie},
  journal={arXiv preprint arXiv:2605.12728},
  year={2026}
}

@inproceedings{robinson2006counting,
  title={Counting unlabeled acyclic digraphs},
  author={Robinson, Robert W},
  booktitle={Combinatorial Mathematics V: Proceedings of the Fifth Australian Conference, Held at the Royal Melbourne Institute of Technology, August 24--26, 1976},
  pages={28--43},
  year={2006},
  organization={Springer}
}

@article{liu2026toolgate,
  title={ToolGate: Contract-Grounded and Verified Tool Execution for LLMs},
  author={Liu, Yanming and Peng, Xinyue and Cao, Jiannan and Wang, Xinyi and Deng, Songhang and Chen, Jintao and Yin, Jianwei and Zhang, Xuhong},
  journal={arXiv preprint arXiv:2601.04688},
  year={2026}
}

@article{BADMUS2027113555,
title = {PowerChain: A verifiable agentic AI system for automating distribution grid analyses},
journal = {Electric Power Systems Research},
volume = {262},
pages = {113555},
year = {2027},
issn = {0378-7796},
doi = {https://doi.org/10.1016/j.epsr.2026.113555},
url = {https://www.sciencedirect.com/science/article/pii/S0378779626008485},
author = {Emmanuel O. Badmus and Peng Sang and Dimitrios Stamoulis and Amritanshu Pandey}
}

@inproceedings{jin2024chatgrid,
  title={ChatGrid: Power grid visualization empowered by a large language model},
  author={Jin, Sichen and Abhyankar, Shrirang},
  booktitle={2024 IEEE Workshop on Energy Data Visualization (EnergyVis)},
  pages={12--17},
  year={2024},
  organization={IEEE}
}

@article{wang2025pro2guard,
  title={Pro2Guard: Proactive Runtime Enforcement of LLM Agent Safety via Probabilistic Model Checking},
  author={Wang, Haoyu and Poskitt, Christopher M and Sun, Jun and Wei, Jiali},
  journal={arXiv preprint arXiv:2508.00500},
  year={2025}
}

@article{jia2025enhancing,
  title={Enhancing LLMs for power system simulations: A feedback-driven multi-agent framework},
  author={Jia, Mengshuo and Cui, Zeyu and Hug, Gabriela},
  journal={IEEE Transactions on Smart Grid},
  year={2025},
  publisher={IEEE}
}

@book{manning2008introduction,
  title={Introduction to Information Retrieval},
  author={Manning, Christopher D and Raghavan, Prabhakar and Sch{\"u}tze, Hinrich},
  year={2008},
  publisher={Cambridge University Press}
}

@article{jarvelin2002cumulated,
  title={Cumulated gain-based evaluation of {IR} techniques},
  author={J{\"a}rvelin, Kalervo and Kek{\"a}l{\"a}inen, Jaana},
  journal={ACM Transactions on Information Systems},
  volume={20},
  number={4},
  pages={422--446},
  year={2002},
  publisher={ACM},
  doi={10.1145/582415.582418}
}

\end{document}